%% file: paper.tex
\documentclass{tlp}

\usepackage[utf8]{inputenc}
\usepackage{amsmath}
\usepackage{amssymb}
\usepackage{microtype}
\usepackage{url}
\usepackage{listings}
\usepackage{xcolor}
\usepackage{graphicx}

\input{macro}

\begin{document}

\lefttitle{Joachim Baumeister et al.}

\jnlPage{1}{8}
\jnlDoiYr{2021}
\doival{10.1017/xxxxx}

\title[Industrial-scale Product Configuration]{Towards Industrial-scale Product Configuration\thanks{
    This work was partly funded by ZIM BMBK grants KK5291307GR4 and KK5394902GR4.
    We would like to thank Simon Vandevelde and the anonymous reviewers for their valuable feedback.}}
\begin{authgrp}
  \author{\sn{Joachim} \gn{Baumeister}}
  \affiliation{denkbares, W\"urzburg, Germany}\affiliation{University of W\"urzburg, Germany}
  \author{\sn{Susana} \gn{Hahn}}
  \affiliation{University of Potsdam, Germany}\affiliation{Potassco Solutions, Germany}
  \author{\sn{Konstantin} \gn{Herud}}
  \affiliation{denkbares, W\"urzburg, Germany}
  \author{\sn{Max} \gn{Ostrowski}}
  \affiliation{Potassco Solutions, Germany}
  \author{\sn{Jochen} \gn{Reutelsh\"ofer}}
  \affiliation{denkbares, W\"urzburg, Germany}
  \author{\sn{Nicolas} \gn{R\"uhling}}
  \affiliation{University of Potsdam, Germany}\affiliation{UP Transfer, Germany}
  \author{\sn{Torsten} \gn{Schaub}}
  \affiliation{University of Potsdam, Germany}\affiliation{Potassco Solutions, Germany}
  \author{\sn{Philipp} \gn{Wanko}}
  \affiliation{Potassco Solutions, Germany}
\end{authgrp}
\maketitle
\begin{abstract}
  We address the challenge of product configuration in the context of increasing customer demand for diverse and complex products.
  We propose a solution through a curated selection of product model benchmarks formulated in the \coom\ language,
  divided into three fragments of increasing complexity.
  Each fragment is accompanied by a corresponding example on bike configuration,
  and additional scalable product models are included in the \coomsuite,
  along with relevant resources.
  We outline an ASP-based workflow for solving \coom-based configuration problems,
  highlighting its adaptability to different paradigms and alternative ASP solutions.
  The \coomsuite\ aims to provide a comprehensive, accessible, and representative set of examples that can serve
  as a common ground for stakeholders in the field of product configuration.
\end{abstract}


\section{Introduction}

Configuration~\citep{kbc14}, a longstanding challenge in AI,
is gaining renewed attention as manufacturing industries grapple with the
challenge of meeting customer demands for high variance and complexity,
all while operating under a high level of automation.
To achieve this, configuration processes must empower customers to tailor products to their specific needs within a predefined configuration space.
This space is designed to seamlessly integrate with subsequent processes (pricing, quotation, manufacturing, resource planning, and delivery),
enabling them to be executed in a highly automated manner.
Despite the existence of various approaches to product configuration~\citep{kbc14},
building and utilizing complex product models for this purpose remains a challenge in industrial settings
due to the diversity of methods, representations, and configuration systems available.
Furthermore, stakeholders from industry, research, and software development often struggle
to align their understanding of the practical application task with the precise semantics of the modeling formalism.

To address this challenge,
we propose a curated selection of product model benchmarks that encapsulate the key challenges of product configuration.
These benchmarks can serve as a common ground for fostering a meaningful exchange among all stakeholders involved.
More precisely,
we formulate our benchmarks in the industrial configuration language \coom~\citep{coomlang},%
\footnote{\Coom\ is developed by denkbares and used in numerous industrial applications.}
and distinguish three language fragments of increasing complexity.
The basic one, \core, is mainly about (discrete) attributes and resembles a simple constraint satisfaction problem.
The extended one, \poom, adds partonomies and cardinalities.
And the last considered \coom\ fragment, \xoom\ adds numeric variables and calculations on top of \poom.
Each such class is accompanied by an exemplary product model from the domain of bikes,
more precisely, a \kidsbike, \citybike, and \travelbike\ model in \coom.
Extensions for two of these language fragments called \openpoom\ and \openxoom, respectively, enable modeling with unbounded cardinalities,
where the latter is exemplified by the product model of a \cargobike.
These and further scalable product models are bundled together in the \coomsuite~\citep{coomsuite},
which is intended to serve as a workbench for experimentation with industrial-scale product configuration problems, and
in addition to all benchmark files, contains \coom\ grammar definitions, and further resources related to Answer Set Programming (ASP; \citealp{lifschitz19a}).

\begin{figure}
  \centering
  \includegraphics[width=\textwidth]{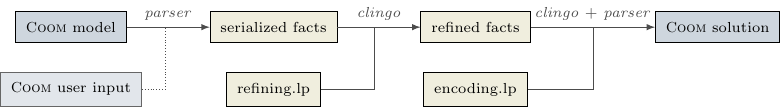}
  \caption{Workflow for solving \coom\ configuration problems with ASP}
  \label{fig:workflow}
\end{figure}

The workflow for solving \coom-based configuration problems with ASP is outlined in Figure~\ref{fig:workflow}.
We start by converting product models in the form of a \coom\ specification and, optionally, a user input into ASP facts.
This conversion is guided by a \coom\ grammar, which is supplied through the \coomsuite.
A specialized ASP visitor then processes the resulting parse tree,
translating it into a corresponding set of facts.
This initial set of facts serves as a direct serialization of the tree structure and
is further refined through an ASP encoding tailored to the specific ASP solution being implemented.
The refined fact format, in conjunction with its corresponding configuration encoding, comprehensively defines the configuration space.
Ultimately, the answer sets representing valid configurations are transformed back into \coom\ solutions.
%
%
Additionally, the \coomsuite\ offers the possibility to interactively solve configuration problems through a \coom-specific
but application-independent user interface (UI).

While this workflow is primarily geared towards ASP,
it is adaptable in several ways.
First, the serialization of \coom\ specifications can be tailored for different paradigms by utilizing a specialized parse tree visitor.
Second, alternative ASP solutions can be investigated by modifying the refinement of the fact format and
employing a different configuration encoding.

Our goals in developing the \coomsuite\ are to provide and evaluate a comprehensive set of product model examples that are:
\begin{description}
  \item [Accessible] Easily understandable for individuals with diverse backgrounds, including researchers and product managers.
  \item [Representative] Encompass the typical challenges encountered in product configuration over recent decades.
  \item [Unifying] Serve as a shared reference point for discussions on representations and formalisms in the field.
  \item [Facilitative] Aid in the development, testing, and comparison of software components designed for product configuration.
  \item [Open] Freely available for public use and contribution.
\end{description}
To the best of our knowledge,
there currently exists no publicly accessible set of product configuration examples that meets our criteria.

The remainder of our paper is organized as follows:
Section~\ref{sec:core} introduces the \core\ language and demonstrates its application through the \kidsbike\ example.
Section~\ref{sec:asp} presents our ASP-based approach to product configuration,
detailing the refined fact format and ASP encoding.
Section~\ref{sec:extensions} outlines various extensions to the \core\ language which contain commonly used features in product modeling
and how they are handled within our ASP framework.
These include concepts such as partonomy, cardinalities, and numerics,
as well as user input and unbounded cardinalities.
Further, an alternative encoding for solver \flingo~\citep{cafascwa26a} is presented.
Finally, Section~\ref{sec:ui} describes the \coomsuite\ UI along with its ASP implementation
through system \clinguin\ and
Section~\ref{sec:benchmarks} demonstrates the practical value of the \coomsuite\ by
evaluating alternative ASP solutions against benchmarks of varying complexity levels.

\subsection*{Related work}\label{sec:related}

Early on, a general ontology for representing product configuration problems was introduced in~\citep{sotimasu98a},
which inspired many of \coom's concepts.
\Coom\ also draws inspiration from other ontologies aiming to represent corporate knowledge
in a standardized way~\citep{brmawazhwopasr07a,coheluolpascst09a},
especially in the context of products and industrial manufacturing~\citep{lifobi97,baumeister18a}.
A wide range of approaches exist for representing and solving configuration problems
across various paradigms~\citep{junker06a,hofestrybawo14a}.
In recent years, ASP has emerged as a promising alternative,
as evidenced by several applications~%
\citep{%
  gekasc11c,%
  fefaateruraz17a,%
  gescer19a,%
  hebasasc22a%
}.
Moreover, \cite{faryscsh15a} developed an object-oriented approach to configuration by directly defining concepts in ASP,
and \cite{ruscst23a} recently proposed an ASP-based approach rooted in firm mathematical foundations.
In the context of interactive configuration,
\cite{fahakrscscta20a}~conducted a comparative evaluation of various systems,
including the ASP solver \clingo\ as well as SAT and CP systems,
for their suitability in this context, finding \clingo\ to be as capable as any other system.

Another system for interactive configuration has been developed by \cite{cavavede23a},
which uses the IDP-Z3 reasoning engine to interactively solve configuration problems specified in the FO$(\cdot)$ language.
This has been successfully applied to various applications such as finding suitable adhesives \citep{vavejodowi24a}
and machine component design \citep{aedeve22a}.

Furthermore, in the related context of product-line variability modeling and configuration,
the Clafer language~\citep{jusamuanczwa19a} has been developed,
which allows to capture in a unified way feature models~\citep{fefabe24a}, component models, and others.
A series of supporting tools~\citep{anbamuollicz13a} are available of which we highlight
the interactive, web-based Clafer Configurator~\citep{claferconf}.

\section{The \core\ language}\label{sec:core}

In this section,
we introduce the \core\ language fragment through the illustrative \kidsbike\ example.
This fragment primarily corresponds to basic constraint satisfaction problems~\citep{dechter03a},
focusing on variables that need to satisfy a given set of constraints.
The \kidsbike\ example, presented in Listing~\ref{lst:kidsbike:coom},
encapsulates a simplified product configuration problem and incorporates all the essential language features of \core.
%
\begin{lstlisting}[float=ht,language=coom,label=lst:kidsbike:coom,caption={Representation of the \kidsbike\ example in the \coom\ language}]
product {                                                |\label{lst:kidsbike:coom:product}|
   Color   color                                         |\label{lst:kidsbike:coom:features:start}|
   Bool    wheelSupport
   Wheel   frontWheel
   Wheel   rearWheel                                     |\label{lst:kidsbike:coom:features:end}|
}
enumeration Color { Red Green Yellow Blue }              |\label{lst:kidsbike:coom:color}|

enumeration Wheel {                                      |\label{lst:kidsbike:coom:wheel}|
   attribute  num  size                                  |\label{lst:kidsbike:coom:wheel:attribute:size}|
   attribute  num  price                                 |\label{lst:kidsbike:coom:wheel:attribute:price}|

   W14 = ( 14 50 )   W16 = ( 16 60 )                     |\label{lst:kidsbike:coom:wheel:options:start}|
   W18 = ( 18 70 )   W20 = ( 20 80 )                     |\label{lst:kidsbike:coom:wheel:options:end}|
}
behavior {                                               |\label{lst:kidsbike:coom:behavior:start}|
   condition color = Yellow require frontWheel.size > 16 |\label{lst:kidsbike:coom:condition}\label{lst:kidsbike:coom:require}|

   combinations  (wheelSupport  rearWheel)               |\label{lst:kidsbike:coom:combinations}|
   allow         (True          (W14, W16))              |\label{lst:kidsbike:coom:allow:start}|
   allow         (False         (W18, W20))              |\label{lst:kidsbike:coom:allow:end}|

   require frontWheel.size = rearWheel.size              |\label{lst:kidsbike:coom:require:wheels}|
}                                                        |\label{lst:kidsbike:coom:behavior:end}|
\end{lstlisting}
%
Line~\ref{lst:kidsbike:coom:product} establishes the definition of a \lstinline{product}.
This represents the fundamental entity of the object we intend to configure.
Note that only one product can be defined within a single \coom\ file.
Inside a \lstinline{product} several \emph{features} are declared
(Lines~\ref{lst:kidsbike:coom:features:start}\nobreakdash-\ref{lst:kidsbike:coom:features:end}).
A feature is declared by stating its type followed by its name, as in `\lstinline{Color color}'.
In this case, type and name are differentiated solely by capitalization.
However, for the type \lstinline{Wheel}, we have two distinct names: \lstinline{frontWheel} and
\lstinline{rearWheel}.
The type \lstinline{Bool} is a built-in type in \coom.
In \core, a type of a feature must refer to an \emph{enumeration},
representing finite domains of predefined choices.
A simple example is the \lstinline{Color} enumeration in Line~\ref{lst:kidsbike:coom:color}.
Furthermore, enumerations can include \emph{attributes} to assign specific values to each option.
For instance, the \lstinline{Wheel} type defines two attributes: \lstinline{size} and \lstinline{price}
(Line~\ref{lst:kidsbike:coom:wheel:attribute:size} and \ref{lst:kidsbike:coom:wheel:attribute:price}),
and its options are declared in
Lines~\ref{lst:kidsbike:coom:wheel:options:start}\nobreakdash-\ref{lst:kidsbike:coom:wheel:options:end}.
Unlike the \lstinline{Color} enumeration, each \lstinline{Wheel} option includes additional values for the \lstinline{size} and \lstinline{price} attributes,
enclosed within parentheses.
This indicates, for example, that wheel \lstinline{W16} has a \lstinline{size} of \lstinline{16} and a \lstinline{price} of \lstinline{60}.

Features in \core\ can be seen as variables and thus be used for expressing constraints over them.
\Coom\ also offers to use attributes of enumerations as variables,
e.g., by writing \lstinline{frontWheel.size} to refer to the \lstinline{size} attribute of the feature \lstinline{frontWheel}.
In general,
these variables are called \emph{path expressions}.

Constraints can be declared within \lstinline{behavior} structures.
Here, we allow for three types of constraints.
In Line~\ref{lst:kidsbike:coom:condition}, a \emph{conditional requirement} is declared,
stating that the \lstinline{size} of the front wheel has to be greater than \lstinline{16}
if the \lstinline{color} of the bike is \lstinline{Yellow}.
Such constraints are declared by using the keywords \lstinline{condition} and \lstinline{require},
followed by a simple logical (comparison) statement.
These statements can contain path expressions and constants, numbers, or strings.
In Line~\ref{lst:kidsbike:coom:require:wheels}, a condition-free requirement is defined,
stating that the size of the front and rear wheel have to be equal.

Another type of constraint is implemented through \emph{combination tables}.
In the example, Lines~\ref{lst:kidsbike:coom:combinations}\nobreakdash-\ref{lst:kidsbike:coom:allow:end}
specify that the use of a \lstinline{wheelSupport} is obligatory with small wheels.
More specifically,
this constraint is declared using the keyword \lstinline{combinations} in Line~\ref{lst:kidsbike:coom:combinations},
followed by a list of path expressions as the table's column headers.
Then, Lines~\ref{lst:kidsbike:coom:allow:start} and~\ref{lst:kidsbike:coom:allow:end} define the allowed combinations.
A table entry can contain multiple values, as illustrated in the \lstinline{rearWheel} column.
Any combination that aligns with an \lstinline{allow} line within the table represents a valid configuration for the product,
while those that do not are invalid.

\section{Solving \Coom-based product configuration with ASP}\label{sec:asp}

We now present our ASP-based approach to solving product configuration problems defined in \core.
As previously mentioned,
we begin by converting a \coom\ input file into a set of facts that represent the serialized parse tree of the \coom\ model.
This is accomplished by using a custom Python \textsl{ANTLR~v4} parser,
which utilizes a \coom\ grammar, making it easily adaptable to future language upgrades.
Next, we transform\footnote{We omit this syntactic translation since it provides no further insights.} the initial set of syntax-reflecting facts into a more refined representation that
captures the essential concepts of the configuration problem in view of the configuration encoding at hand.
This abstraction allows us to move beyond the specific details of the input format and
focus on modeling the fundamental concepts of the configuration problem.

More precisely,
the basic concepts of the refined fact format are variables whose type
is a discrete attribute over a finite domain.
There are two types of constraints over these variables: Boolean and table constraints.
For illustration,
we present (an extract of) the refined fact format of the \kidsbike\ in Listing~\ref{lst:kidsbike:asp}.
Our fact format uses an object-centered representation,
which identifies all concepts with terms.

Variables are identified by strings such as \lstinline{"color"} or \lstinline{"frontWheel"}
and their types are declared in Lines~\ref{lst:kidsbike:asp:type:start}\nobreakdash-\ref{lst:kidsbike:asp:type:end}.
%
\begin{lstlisting}[language=clingos,label=lst:kidsbike:asp,linerange={1-3,6-22},caption={Extract of the refined ASP representation of the \kidsbike\ example}]
type("color","Color").      type("wheelSupport","Bool").                %%#(\label{lst:kidsbike:asp:type:start}#)
type("frontWheel","Wheel"). type("frontWheel.size","size").
type("rearWheel","Wheel").  type("rearWheel.size","size").              %%#(\label{lst:kidsbike:asp:type:end}#)

discrete("Wheel").  domain("Wheel",("W14";"W16";"W18";"W20")).          %%#(\label{lst:kidsbike:asp:attr:wheel}#)
discrete("size").   domain("size",(14;16;18;20)).                       %%#(\label{lst:kidsbike:asp:attr:size}#)
discrete("Bool").   domain("Bool",("True";"False")).                    %%#(\label{lst:kidsbike:asp:attr:bool}#)

constraint(c0,"boolean").                                               %%#(\label{lst:kidsbike:asp:constraint:0:start}#)
binary(c0,f0,"||",f1).                 unary(f0,"!",f2).
binary(f2,"color","=","Yellow").       constant("Yellow").
binary(f1,"frontWheel.size",">","16"). number("16",16).                 %%#(\label{lst:kidsbike:asp:constraint:0:end}#)

constraint(c1,"table").                                                 %%#(\label{lst:kidsbike:asp:constraint:1:start}#)
column(c1,0,"wheelSupport"). column(c1,1,"rearWheel").                  %%#(\label{lst:kidsbike:asp:constraint:1:column}#)
allow(c1,(0,0),"True").      allow(c1,(1,0),("W14";"W16")).             %%#(\label{lst:kidsbike:asp:constraint:1:allow:start}#)
allow(c1,(0,1),"False").     allow(c1,(1,1),("W18";"W20")).             %%#(\label{lst:kidsbike:asp:constraint:1:end}#)

constraint(c2,"boolean").                                               %%#(\label{lst:kidsbike:asp:constraint:2:start}#)
binary(c2,"frontWheel.size","=","rearWheel.size").                      %%#(\label{lst:kidsbike:asp:constraint:2:end}#)
\end{lstlisting}

Lines~\ref{lst:kidsbike:asp:attr:wheel}\nobreakdash-\ref{lst:kidsbike:asp:attr:bool} give
the (discrete) attributes of the configuration model.
In fact,
both the \lstinline{Wheel} enumeration and one of its enumeration attributes \lstinline{size} from the \coom\ model
are represented as discrete attributes.
The \lstinline{Bool} attribute is a \coom\ built-in and added automatically during refinement.
We omit showing the \lstinline{Color} and \lstinline{price} attribute.
In \coom, the (combined) values of such attributes are among the enumeration's options (cf.~Section~\ref{sec:core}).
We guarantee their compatibility, e.g., of \lstinline{size} and \lstinline{price} from the \lstinline{Wheel} enumeration, by means of a table constraint which is omitted for brevity.

Lines~\ref{lst:kidsbike:asp:constraint:0:start}\nobreakdash-\ref{lst:kidsbike:asp:constraint:2:end}
give three constraints of the configuration model;
they reflect \coom\ \lstinline{behavior}[s].
Each constraint is identified with a term, \lstinline{c0}, \lstinline{c1}, etc., during refinement.
Constraint \lstinline{c0} is a conditional requirement stating that the color \lstinline{Yellow} is only available with larger wheels.
For this, we require in Lines~\ref{lst:kidsbike:asp:constraint:0:start}\nobreakdash-\ref{lst:kidsbike:asp:constraint:0:end}
that \lstinline{color=Yellow} is false or \lstinline{frontWheel.size>16} is true.
These statements are represented by \lstinline{f0} and \lstinline{f1}, respectively,
and the operators \lstinline{"||"} and \lstinline{"!"} stand for
disjunction and negation, respectively.
They are further decomposed via predicates \lstinline{binary/4} and \lstinline{unary/3}
until the variable or constant level is reached.
Analogously,
Lines~\ref{lst:kidsbike:asp:constraint:2:start}\nobreakdash-\ref{lst:kidsbike:asp:constraint:2:end} encode a Boolean constraint
stating that the size of the front and rear wheel must be equal.

Lastly,
Lines~\ref{lst:kidsbike:asp:constraint:1:start}\nobreakdash-\ref{lst:kidsbike:asp:constraint:1:end} give a table constraint
reflecting a \lstinline{combinations} table in \coom.
The constraint and its columns are declared in Lines~\ref{lst:kidsbike:asp:constraint:1:start}
and~\ref{lst:kidsbike:asp:constraint:1:column},
followed by the table entries in
Lines~\ref{lst:kidsbike:asp:constraint:1:allow:start} and~\ref{lst:kidsbike:asp:constraint:1:end},
expressed by predicate \lstinline{allow/3}.
Note that multiple values can be given for a table entry, as for entries \lstinline{(1,0)} and \lstinline{(1,1)}.

In what follows,
we present a simple configuration encoding for \core\ models in Listing~\ref{lst:encoding:core}.
While Line~\ref{lst:encoding:core:discrete} generates exactly one value per discrete attribute from an associated domain,
Line~\ref{lst:encoding:core:aux} generates auxiliary values for constants used in the constraints.
%
\begin{lstlisting}[language=clingos,label=lst:encoding:core,caption={ASP encoding of refined \core}]
{ value(X,V) : domain(T,V) } = 1 :- type(X,T), discrete(T).     %%#(\label{lst:encoding:core:discrete}#)
value(P,P) :- constant(P). value(P,N) :- number(P,N).           %%#(\label{lst:encoding:core:aux}#)

:- constraint(C,_), not satisfied(C).                           %%#(\label{lst:encoding:core:constraint:violate}#)

satisfied(F) :- binary(F,XL,"||",XR),                           %%#(\label{lst:encoding:core:constraint:disj:start}#)
                1 <= { satisfied(XL); satisfied(XR) }.          %%#(\label{lst:encoding:core:constraint:disj:end}#)
satisfied(F) :- unary(F,"!",F'), not satisfied(F').             %%#(\label{lst:encoding:core:constraint:neg}#)

satisfied(F) :- binary(F,XL,"=",XR), VL = VR,                   %%#(\label{lst:encoding:core:constraint:eq:start}#)
                value(XL,VL), value(XR,VR).                     %%#(\label{lst:encoding:core:constraint:eq:end}#)
satisfied(F) :- binary(F,XL,">",XR), VL > VR,                   %%#(\label{lst:encoding:core:constraint:gr:start}#)
                value(XL,VL), value(XR,VR).                     %%#(\label{lst:encoding:core:constraint:gr:end}#)

nhit(C,Row)  :- allow(C,(Col,Row),_), column(C,Col,X),          %%#(\label{lst:encoding:core:constraint:nhitrow:start}#)
                not value(X,V) : allow(C,(Col,Row),V).          %%#(\label{lst:encoding:core:constraint:nhitrow:end}#)
satisfied(C) :- allow(C,(_,Row),_), not nhit(C,Row).            %%#(\label{lst:encoding:core:constraint:table}#)

#show value/2.                                                  %%#(\label{lst:encoding:core:show}#)

\end{lstlisting}

Next, Lines~\ref{lst:encoding:core:constraint:violate}\nobreakdash-\ref{lst:encoding:core:constraint:table}
encode Boolean and table constraints.
To start with, the integrity constraint in Line~\ref{lst:encoding:core:constraint:violate} makes sure that all
original constraints are satisfied.
Lines~\ref{lst:encoding:core:constraint:disj:start}\nobreakdash-\ref{lst:encoding:core:constraint:gr:end} specify
satisfaction conditions for various types of binary and unary formulas making up a Boolean constraint, viz.\
disjunction (Lines~\ref{lst:encoding:core:constraint:disj:start} and \ref{lst:encoding:core:constraint:disj:end}),
negation (Line~\ref{lst:encoding:core:constraint:neg}), and
comparison operators \lstinline{=} and \lstinline{>} (Lines~\ref{lst:encoding:core:constraint:eq:start}\nobreakdash-\ref{lst:encoding:core:constraint:gr:end}).
To encode disjunction,
we make use of \clingo's weight constraints,
where the expression \lstinline|1 <= { satisfied(XL); satisfied(XR) }|
states that at least one of the two subformulas \lstinline{XL} or \lstinline{XR} must be satisfied.

Lines~\ref{lst:encoding:core:constraint:nhitrow:start}\nobreakdash-\ref{lst:encoding:core:constraint:table} give
satisfaction conditions for table constraints.
For this,
we identify all rows that are not satisfied (Lines~\ref{lst:encoding:core:constraint:nhitrow:start} and~\ref{lst:encoding:core:constraint:nhitrow:end}).
A row is not satisfied whenever one of its entries is not satisfied.
Note that as entries can contain multiple values,
we employ \clingo's conditional literals to check for all of the individual values within the table entry.
Here, the expression \lstinline{not value(X,V) : allow(C,(Col,Row),V)}
will be expanded to \lstinline|not value(X,V1), not value(X,V2),...| during grounding
for a table entry containing the tuple \lstinline{(V1,V2,...)}.
More precisely, the expression \lstinline{not value(X,V) : allow(C,(Col,Row),V)} gathers all possible values \lstinline{V1,V2,...} for a given table entry \lstinline{(Col,Row)} of the table constraint \lstinline{C}
and then expands to a conjunction of literals \lstinline{not value(X,V1), not value(X,V2),...}.
Finally, a table constraint is satisfied if one of its rows is satisfied (Line~\ref{lst:encoding:core:constraint:table}).

Listing~\ref{lst:kidsbike:cli} shows how to solve the \kidsbike\ example with the \lstinline{solve} mode of the \coomsuite\
following the workflow from Figure~\ref{fig:workflow}.
Currently, the \coomsuite\ uses \clingo\ as the default solver.
Here, \lstinline{kids-bike.coom} is the input file containing the \coom\ model of the \kidsbike\ example
corresponding to Listing~\ref{lst:kidsbike:coom}.
The option \lstinline{--output coom} (or \lstinline{-o coom} for short) converts the ASP output to a (more readable) \coom\ format.
%
\begin{lstlisting}[language=shell,label=lst:kidsbike:cli,caption={Solving the \kidsbike\ example with the \coomsuite.}]
$ coomsuite solve kids-bike.coom --output coom
COOM Suite version 0.1
Reading from kids-bike.lp
Solving...
Answer: 1
color[0] = "Red"
wheelSupport[0] = "True"
frontWheel[0] = "W14"   frontWheel[0].size[0] = 14
rearWheel[0]  = "W14"   rearWheel[0].size[0]  = 14
SATISFIABLE
\end{lstlisting}

\section{\Coom\ language extensions}\label{sec:extensions}

Having explained the basics of \coom,
we now proceed to elaborate on its extensions
and how they can be modelled and solved using ASP within our framework.

We begin by showing how more commonly used features
like partonomy, cardinalities, and numeric calculations can be
modeled within \xoom, using our \travelbike\ example for illustration.
We do not delve into \poom, as all its concepts are already encompassed within \xoom.

We start by explaining how these features are implemented
and how they are used in the \travelbike\ example.
The bike's partonomy consists of two parts: a carrier and a frame.
Both of them can be equipped with zero, one, or multiple bags, thus making bags an optional component.
However, the model's constraints can necessitate a minimum total storage volume greater than zero,
effectively requiring the inclusion of at least one bag.
To calculate the total storage volume, arithmetic aggregation functions are employed
to sum the storage capacity across all bags.

Similar to above, we first outline how these features are modeled within \xoom\ and then
demonstrate their representation and resolution using ASP.
For this part, we first present the fact format and then proceed to show two similar but alternative encodings:
One for solver \clingo\ and one for \flingo.
For both, we showcase selected aspects and refer the reader to~\cite{coomsuite} for the complete encoding.

In the second part of this section, we use \xoom\ as a basis to describe two less commonly used features
which are nevertheless equally important in practice.
The first one is so-called user input which allows for the specification of custom requirements at runtime
such as for example a minimum total storage volume for the abovementioned \travelbike,
thereby excluding certain solutions from the search space beforehand.

The second feature are unbounded cardinalities, which allow modeling cases
where the exact (or approximate) number of some object is not known beforehand.
Here, we introduce extensions to two of our language fragments, which we call \openpoom\ and \openxoom, respectively.
We illustrate \openxoom\ using the \cargobike\ example, a modified version of the \travelbike, where the number of possible bags is unbounded.
For both, we proceed in the same manner as before by first showcasing a \coom\ example and then
showing how this is solved using ASP.

\subsection{The \travelbike\ in \xoom}\label{sec:extensions:coom}
\begin{lstlisting}[float=ht,language=coom,label=lst:travelbike:coom,caption={Simplified representation of the \travelbike\ in \coom}]
product {                                               |\label{lst:travelbike:coom:product:start}|
    num 0-200 totalVolume                               |\label{lst:travelbike:coom:product:totalvol}|
    num 0-200 requestedVolume                           |\label{lst:travelbike:coom:product:reqvol}|
    Carrier   carrier                                   |\label{lst:travelbike:coom:product:carrier}|
    Frame     frame                                     |\label{lst:travelbike:coom:product:frame}|
}                                                       |\label{lst:travelbike:coom:product:end}|
structure Carrier {0..3 Bag bag}                        |\label{lst:travelbike:coom:carrier}|
structure Frame   {0..2 Bag bag}                        |\label{lst:travelbike:coom:frame}|

enumeration Bag {                                       |\label{lst:travelbike:coom:bag:start}|
    attribute num volume                                |\label{lst:travelbike:coom:bag:volume}|
    B20  = ( 20 )   B50  = ( 50 )   B100 = ( 100 )       |\label{lst:travelbike:coom:bag:options}|
}                                                       |\label{lst:travelbike:coom:bag:end}|
behavior {
    require count(carrier.bag) + count(frame.bag) <= 4  |\label{lst:travelbike:coom:constraint:count}|

    require totalVolume = sum(carrier.bag.volume) +     |\label{lst:travelbike:coom:constraint:sum:start}|
                          sum(frame.bag.volume)         |\label{lst:travelbike:coom:constraint:sum:end}|

    require totalVolume >= requestedVolume              |\label{lst:travelbike:coom:constraint:volume}|
}
\end{lstlisting}
%
Listing~\ref{lst:travelbike:coom} gives a simplified \xoom\ representation of the \travelbike\ example.
First, the features of the bike are defined,
starting with the two numeric ones, \lstinline{totalVolume} and \lstinline{requestedVolume}
in Lines~\ref{lst:travelbike:coom:product:totalvol} and~\ref{lst:travelbike:coom:product:reqvol}.
Both are marked as such by the keyword \lstinline{num} at the beginning of the line,
followed by their respective ranges.
The second feature can be thought of as a user requirement that can be set at runtime (cf.~Section~\ref{sec:userinput}).
Next, Lines~\ref{lst:travelbike:coom:product:carrier} and~\ref{lst:travelbike:coom:product:frame}
define the \lstinline{carrier} and \lstinline{frame},
whose types are captured by \lstinline{structure}[s]
in Lines~\ref{lst:travelbike:coom:carrier} and \ref{lst:travelbike:coom:frame}.
Both the \lstinline{num} keyword and the \lstinline{structure} keyword
were not possible before in \core.
Other than \lstinline{enumeration}[s],
\lstinline{structure}[s] may have features on their own,
which allows for building complex partonomies.
For example,
the \lstinline{Carrier} and the \lstinline{Frame} {structure} have exactly one feature, \lstinline{bag}.
The expressions \lstinline{0..3} and \lstinline{0..2} give their respective cardinalities.
Strictly speaking, each feature in \coom\ has a cardinality, but when omitted it defaults to~\lstinline{1..1}.
In all our \coom\ fragments until now, lower and upper bounds are mandatory,
and we carry this requirement over to our fact format as well
(only in Section~\ref{sec:open} we generalize this to allow for open bounds.)
The type of both features is \lstinline{Bag},
which in turn is an \lstinline{enumeration} with a single attribute \lstinline{volume}
(Lines~\ref{lst:travelbike:coom:bag:start}\nobreakdash-\ref{lst:travelbike:coom:bag:end}).

Lastly, we discuss the constraints of the model.
The first constraint requires that there are no more than four bags in a configuration
(Line~\ref{lst:travelbike:coom:constraint:count}).
For this,
it uses aggregate functions to \lstinline{count} the number of bags.
In general, aggregate functions perform calculations over a set of variables,
which are defined implicitly in terms of a path expression, e.g., \lstinline{carrier.bag}.
In this case, the function \lstinline{count(carrier.bag)} returns the actual number of bags
attached to the \lstinline{carrier} in the configuration at hand.

We introduced path expressions in Section~\ref{sec:core} as variables corresponding to a product feature, or
an enumeration attribute, respectively.
This is valid for \core, however, for \poom\ and \xoom\ we need a more general definition as features can have cardinalities different than~1.
We now say that a path expression serves as an identifier for a set of variables.%
\footnote{In \core\ this reduces to the case of singleton sets with cardinality~1.}
Each part of the path expression is a feature name and the last part can be an enumeration attribute.

Another aggregate function is \lstinline{sum},
which returns the sum of the values of all variables in a set
(Lines~\ref{lst:travelbike:coom:constraint:sum:start} and \ref{lst:travelbike:coom:constraint:sum:end}).
Here, the constraint requires that the value of \lstinline{totalVolume} is equal to the sum of the volume of all bags.
The last constraint is another \lstinline{requirement} and
relates the calculated value of \lstinline{totalVolume} to
the value \lstinline{requestedVolume}.

\subsection{Solving \xoom\ in ASP}\label{sec:extensions:asp}

In this section, we outline our ASP-based approach to solving product configuration problems specified in \xoom.
We start by illustrating the refined fact format representation using the \travelbike\ example,
which incorporates more advanced concepts than the fact format presented in Section~\ref{sec:asp}.

The essential concepts of this new fact format are a partonomy with bounded cardinalities,
attributes ranging over discrete or integer domains, and Boolean as well as table constraints
where the former now allow for arithmetic expressions and aggregate functions.

We represent the configuration as a tree such that each node represents a variable,
and its root reflects the object to be configured.
Variables can be either parts or attributes.
Solutions are represented as subtrees of this configuration tree
where values are assigned to any attribute variable included in the subtree.
Thus, not all variables are necessarily included in the solution and an excluded variable renders all variables in its (possible) subtree excluded.
Cardinalities are represented as constraints over sets of nodes in the tree.

\begin{figure}[ht]
  \centering
  \includegraphics[width=\textwidth]{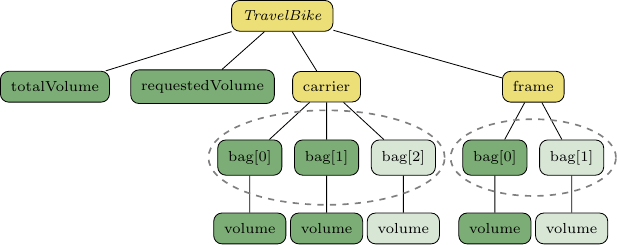}
  \caption{The \travelbike\ example converted into a configuration tree}
  \label{fig:tree}
\end{figure}

\subsubsection{Fact format}\label{sec:extensions:asp:facts}
In \coom, a partonomy is established by defining structures and linking them through features.
These features can also reference enumerations or numeric variables,
potentially with cardinalities (as illustrated in Listing~\ref{lst:travelbike:coom}).
We simplify this approach by treating both part and attribute variables uniformly
and refer to the resulting structure as the \emph{configuration tree}
which can be considered an explicit representation of the \coom\ configuration product model from before.

For the \travelbike, this structure, which serves as the basis for our fact format, is visualized in Figure~\ref{fig:tree}.
Here, nodes belonging to parts are highlighted in yellow and those belonging to attributes in green.
Consider an example solution where the third bag of the carrier and the second bag of the frame
are undefined, thus not included in the solution.
We represent this in Figure~\ref{fig:tree} by highlighting the corresponding variables
in a lighter color.
Note that undefined variables automatically render their subnodes undefined as well.
Note that for the sake of readability variable names are abbreviated and value assignments are omitted.
Cardinalities of features are treated by grouping variables belonging to the same feature
and with the same parent variable in sets (represented by dashed circles in the diagram).

Our refinement process generates all possible variables within the model and
assigns them unique (human-readable) identifiers by leveraging the tree-like structure of the configuration.
For example, the \lstinline{frame} feature of the \lstinline{product} in Listing~\ref{lst:travelbike:coom}
results in the variable \lstinline{"root.frame[0]"}.
In this context, \lstinline{"root"} refers to the object being configured, in our case, the \travelbike.
Indices are required as feature cardinalities are potentially larger than~1.
For instance,
the \lstinline{bag} feature of the \lstinline{Frame} \lstinline{structure}
yields two variables \lstinline{"root.frame[0].bag[0]"} and \lstinline{"root.frame[0].bag[1]"}
which can be read as ``the first and second bag of the first frame'', respectively.
For capturing each such variable in an object-centered representation,
we include three facts representing the type, index and parent of the variable.
\begin{lstlisting}[language=clingos,label=lst:travelbike:asp:vars,linerange=37-39,caption={A variable of the instantiated \travelbike}]
type("root.frame[0].bag[1]","Bag").
index("root.frame[0].bag[1]",1).
parent("root.frame[0].bag[1]","root.frame[0]").
\end{lstlisting}

We represent \coom's \lstinline{structure}[s] as parts.
Hence, we encode the \travelbike's parts as \lstinline{part("Carrier").} and \lstinline{part("Frame").}
Since \lstinline{Bag} is an \lstinline{enumeration} in \coom, it is represented as a discrete attribute
(cf.~Section~\ref{sec:asp}).

Cardinalities are represented as lower bound constraints.
The upper bound is compiled away during our refinement process by generating the corresponding number of variables.
For example, the cardinalities for the bags of the frame are encoded as follows:
%
\begin{lstlisting}[language=clingos,label=lst:travelbike:asp:cardinality,linerange=134-136,caption={Cardinality constraint for the bags of the \travelbike\ frame}]
constraint(("root.frame[0].bag",0),"lowerbound").
set("root.frame[0].bag","root.frame[0].bag[0]").
set("root.frame[0].bag","root.frame[0].bag[1]").
\end{lstlisting}
%
While the second argument \lstinline{"lowerbound"} of predicate \lstinline{constraint/2} marks its type,
the first one contains a pair consisting of
an identifier \lstinline{"root.frame[0].bag"} and the actual lower bound 0.
The identifier acts as a representative for a set of variables.
This set is encoded by means of predicate \lstinline{set/2},
whose first argument identifies the set and its second one a set member.
The upper bound~2 is thus reflected by the number of set elements.

A numeric attribute and its corresponding range is declared as follows:
\begin{lstlisting}[language=clingos,label=lst:travelbike:asp:num,linerange=3-4,caption={Integer attribute \lstinline{totalVolume} of the \travelbike}]
integer("totalVolume").
range("totalVolume",0,200).
\end{lstlisting}
As before, a variable is associated with this attribute through predicate \lstinline{type/2},
e.g., \lstinline{type("root.totalVolume[0]","totalVolume")}.
Currently, the only numeric attribute type is \emph{integer}.

Aggregate functions are represented in analogy to cardinalities in a set-based fashion.
%
\begin{lstlisting}[language=clingos,label=lst:travelbike:asp:count,linerange=76-78,caption={Count aggregate function of the \travelbike}]
function("count(root.frame.bag)","count","root.frame.bag").
set("root.frame.bag","root.frame[0].bag[0]").
set("root.frame.bag","root.frame[0].bag[1]").
\end{lstlisting}
%
The predicate \lstinline{function/3} comprises
the function's identifier,
its type (e.g., \lstinline{count}), and
a set identifier.
As above, the members of the set are declared via \lstinline{set/2}.

Lastly, (binary) arithmetic functions like \lstinline{+} or \lstinline{-} are represented in the same manner as Boolean binary functions (cf.~Listing~\ref{lst:kidsbike:asp}).

\subsubsection{Encoding}\label{sec:extensions:asp:encoding}
We now describe selected aspects of our encoding for solving configuration problems in \xoom.
We start by describing the rules related to the configuration tree.
\begin{lstlisting}[language=clingos,label=lst:encoding:x:include,linerange={1-9},caption={Definedness of variables in \xoom}]
{ include(X) : type(X,_) }. include("root").                    %%#(\label{lst:encoding:x:include:gen}#)

:- include(X), parent(X,P), not include(P).                     %%#(\label{lst:encoding:x:include:parent}#)

:-  include(X),  set(P,X ), index(X, I), I > 0,                %%#(\label{lst:encoding:x:include:index:start}#)
not include(X'), set(P,X'), index(X',I-1).                     %%#(\label{lst:encoding:x:include:index:end}#)

:- constraint((P,L),"lowerbound"), set(P,X), index(X,L-1),     %%#(\label{lst:encoding:x:include:cardinality:start}#)
   parent(X,X'), include(X'), not include(X).                   %%#(\label{lst:encoding:x:include:cardinality:end}#)
\end{lstlisting}
Line~\ref{lst:encoding:x:include:gen} of Listing~\ref{lst:encoding:x:include} generates instances of \lstinline{include/1} predicates,
stating that a variable of the configuration tree is included in the solution.
This rule applies to part as well as attribute variables;
the root variable is always included.
Line~\ref{lst:encoding:x:include:parent} restricts this to variables whose parent variables are included as well.
For symmetry breaking,
Lines~\ref{lst:encoding:x:include:index:start} and \ref{lst:encoding:x:include:index:end}
enforce that a variable with index \lstinline{I} is only included if the variable with index \lstinline{I-1}
from the same (cardinality) set is included as well.
We leverage this for enforcing cardinality constraints in Lines~\ref{lst:encoding:x:include:cardinality:start} and \ref{lst:encoding:x:include:cardinality:end}.
Since variables in a cardinality set are included in ascending index order,
we only need to check that the variable whose index corresponds to the lower bound is included in the solution.
Also, a cardinality is only enforced if the parent variable is included in the solution.

Next, we consider numerics in Listing~\ref{lst:encoding:x:attr} for generating attribute values.
%
\begin{lstlisting}[language=clingos,label=lst:encoding:x:attr,linerange={13-16},caption={Generation of attribute values in \xoom}]
{ value(X,V) : domain(T,V) } = 1 :- include(X), type(X,T),      %%#(\label{lst:encoding:x:attr:discrete:start}#)
                                    discrete(T).                %%#(\label{lst:encoding:x:attr:discrete:end}#)
{ value(X,V) : V = L..U    } = 1 :- include(X), type(X,T),      %%#(\label{lst:encoding:x:attr:integer:start}#)
                                    integer(T), range(T,L,U).   %%#(\label{lst:encoding:x:attr:integer:end}#)
\end{lstlisting}
%
Unlike Section~\ref{sec:asp},
we now need to take into account possibly undefined (attribute) variables.
In our current encoding, we say that an attribute variables is defined iff it is included in the solution.
The rule in Lines~\ref{lst:encoding:x:attr:discrete:start} and \ref{lst:encoding:x:attr:discrete:end} is similar to the one above,
just that now the body additionally contains the atom \lstinline{include(X)}.
This guarantees that an attribute variable only generates a value, if it is included in the solution, and thus, not undefined.
Integer attributes are handled analogously except that values are generated from a range of numbers.

The possible undefinedness of variables also has an effect on constraint satisfaction and violation.
For example, in our \core-encoding, a constraint is violated when it is not satisfied.
This condition is no longer sufficient here, as variables in a constraint can now be undefined.
We therefore need to adapt the integrity constraints accordingly.
%
\begin{lstlisting}[language=clingos,label=lst:encoding:x:cons,linerange={21-23},caption={Boolean constraint checking in \xoom}]
:- constraint((C,F),"boolean"), defined(F), not satisfied(F).
defined(X) :- value(X,_).
defined(F) :- binary(F,XL,_,XR), defined(XL), defined(XR).
\end{lstlisting}
%
The integrity constraint in Listing~\ref{lst:encoding:x:cons} contains the additional atom \lstinline{defined(F)},
making sure that the Boolean formula \lstinline{F} can only be violated if it is actually defined.
The predicate \lstinline{defined/1} is specified recursively.
An (attribute) variable is defined when it has an assigned value and
a binary predicate is defined when both of its parts are defined.

Lastly, we showcase the encoding of arithmetic and aggregate functions.
\begin{lstlisting}[language=clingos,label=lst:encoding:x:num,linerange={36-41},caption={Numerical calculations in \xoom}]
value(F,V) :- function(F,"count",P),
              V = #count{ X    : set(P,X), include(X) }.
value(F,V) :- function(F,"sum",  P),
              V = #sum  { V',X : set(P,X), value(X,V')}.
value(F,VL+VR) :- binary(F,XL,"+",XR),
                  value(XL,VL), value(XR,VR).
\end{lstlisting}
Listing~\ref{lst:encoding:x:num} contains rules for the \lstinline{count} and \lstinline{sum} aggregate function
as well as for the \lstinline{+} operator.
For all three rules we are making use of built-in \clingo\ functionality.
Note that whenever non-domain predicates (\lstinline{include/1} and \lstinline{value/2}) occur inside an aggregate,
assigning the result of that calculation to another variable
possibly creates very large groundings.
This is due to the fact that all possible value assignments
have to be explicitly represented.
However, in this encoding we opted for clarity and simplicity over performance
and leave optimized versions for future work.

Solving \xoom\ product models with the \coomsuite\ works by using its \lstinline{solve} mode
in the same fashion as already shown in Listing~\ref{lst:kidsbike:cli} above for the \kidsbike\ example.
We do not show the command-line in- and output here but defer this to Section~\ref{sec:userinput}
where we introduce user input.

\subsection{Solving \xoom\ with \flingo}\label{sec:extensions:asp:flingo}
An alternative \xoom\ encoding is given for solver \flingo%
\footnote{\sysfont{fclingo} in earlier versions.}
which is a prototype system for solving conditional linear constraints with integer variables in ASP.
This system is a continuation of the \lctocasp\ system~\citep{cakaossc16a} enhanced with conditional linear constraints
given via the translation in~\citep{cafascwa20a}.
\flingo\ uses CASP solver \clingcon~\citep{bakaossc16a} as a backend, therefore being able to deal with large integer ranges and numerical calculations.
The main difference, however, is that variables in \flingo\ can stay undefined
while in \clingcon\ all variables need to have a value assigned.
As \coom\ allows for optional attributes, this is a crucial feature.
The encoding for \flingo\ is identical in many parts to the \clingo\ encoding from the previous section,
except for rules which treat integer attributes and numerical constraints.

We start by giving a brief overview of the \flingo\ features we are utilizing for this adaptation and then proceed to
highlight the differences between the two encodings.
Our main objective with the \flingo\ encoding is to improve performance for numerical instances by overcoming \clingo's grounding bottleneck.
This is achieved in \flingo\ by allowing for a special type of variables which are not subject to grounding.
These are called \emph{integer variables} and as their name suggests, can take values from the domain of the integers.
Importantly, they are being treated differently than propositional variables.
Further, to be able to properly replace the necessary rules from our earlier encoding,
we require that these integer variables can be used to build linear constraints
while comprising a notion of undefinedness and the possibility to define defaults.
Lastly, we also need to be able to exclude variables from a calculation if they do not fulfill certain conditions, i.e., aggregate functions should allow for conditionality.
\flingo\ provides us with all these features as we see in the following.
Rules in \flingo\ are written in the same way as in \clingo\
but additionally there are a few special theory atoms through which we can define integer variables and build up constraints over them.
Here, we explain only those theory atoms that we are using in our encoding and refer the reader to~\cite{cafascwa26a,flingo} for further documentation and examples.
We proceed by showing snippets of the \flingo\ encoding while highlighting the differences from the \clingo\ encoding from Section~\ref{sec:extensions:asp:encoding}.

First, Line~\ref{lst:encoding:flingo:attr:bounded} in Listing~\ref{lst:encoding:flingo:attr} represents the \flingo\ replacement of
the rule in Lines~\ref{lst:encoding:x:attr:integer:start} and \ref{lst:encoding:x:attr:integer:end} in Listing~\ref{lst:encoding:x:attr}
which is responsible for generating possible integer attribute values.
%
\begin{lstlisting}[language=flingos,label=lst:encoding:flingo:attr,linerange={16-17},caption={Generation of integer attribute values in \flingo}]
&in{L..U} =: X :- include(X), type(X,T), integer(T),     range(T,L,U).  %%#(\label{lst:encoding:flingo:attr:bounded}#)
&sum{X}   =  X :- include(X), type(X,T), integer(T), not range(T,_,_).  %%#(\label{lst:encoding:flingo:attr:unbounded}#)
\end{lstlisting}
%
The head atom \lstinline|&in{L..U} =: X|
assigns a values from the range \lstinline{L..U} to variable \lstinline{X}.
Note that since variable \lstinline{X} appears inside the theory atom,
it is declared as an integer variables,
and therefore not subject to grounding.
The operator \lstinline{=:} specifies that this is an \emph{assignment},
meaning that \lstinline{X} only gets a value assigned if the lower and upper bound of the interval, \lstinline{L} and \lstinline{U}, are defined.
In this case, this would not be necessary as both are constants,
however, the current syntax of \flingo\ only allows to use \lstinline{&in} together with the assignment operator \lstinline{=:}.

Due to the fact that integer variables in \flingo\ are freed from grounding (and unlike in the \clingo\ encoding), it is possible to reason
with unbounded ranges (cf.~Line~\ref{lst:encoding:flingo:attr:unbounded}).
When no range is specified, the atom \lstinline|&sum{X} = X| works similar to a choice rule in ASP
in the sense that it defines variable \lstinline{X}
without constraining its range, thus effectively assigning any possible, integer value to it.

Next, Listing~\ref{lst:encoding:flingo:cons} shows some of the rules for Boolean constraint checking
(compare with Listing~\ref{lst:encoding:x:cons}).
%
\begin{lstlisting}[language=flingos,label=lst:encoding:flingo:cons,linerange={41-45},caption={Boolean constraint checking in \flingo}]
defined(F) :- binary(F,X1,_,X2), &df{X1}, &df{X2}.      %%#(\label{lst:encoding:flingo:cons:defined}#)

satisfied(F) :- binary(F,X1,"=", X2), &sum{X1} =  X2.   %%#(\label{lst:encoding:flingo:cons:satisfied:start}#)
satisfied(F) :- binary(F,X1,">=",X2), &sum{X1} >= X2.
satisfied(F) :- binary(F,X1,"<=",X2), &sum{X1} <= X2.   %%#(\label{lst:encoding:flingo:cons:satisfied:end}#)
\end{lstlisting}
%
In Line~\ref{lst:encoding:flingo:cons:defined} we make use of \flingo's \lstinline{&df} atom to check whether a variable is defined.
Binary formulas are checked for satisfaction in the same way as before with the difference that we are using the
\flingo\ atom \lstinline{&sum}.
Recall that since integer variables are treated in a special way in \flingo,
they may only be accessed in the context of theory atoms (or constraints formed by them as in the case of variable \lstinline{X2}).

Lastly, in Listing~\ref{lst:encoding:flingo:num} we show rules for numerical calculations in \flingo\
which correspond to the rules from Listing~\ref{lst:encoding:x:num} for \clingo.
%
\begin{lstlisting}[language=flingos,label=lst:encoding:flingo:num,linerange={47-49},caption={Numerical calculations in \flingo}]
&sum{ 1,X : set(P,X), include(X) } = F :- function(F,"count",P).    %%#(\label{lst:encoding:flingo:num:count}#)
&sum{   X : set(P,X), include(X) } = F :- function(F,"sum",  P).    %%#(\label{lst:encoding:flingo:num:sum}#)
&sum{ XL;  XR } = F :- binary(F,XL,"+",XR).                         %%#(\label{lst:encoding:flingo:num:plus}#)
\end{lstlisting}
They represent calculations of \lstinline{count} and \lstinline{sum} aggregates, as well as for the \lstinline{+} operator.
While in the earlier encoding a different \clingo\ functionality was used for each rule,
here we use the \flingo\ theory atom \lstinline{&sum} for all calculations.
For the \lstinline{count} function, we utilize the \lstinline{&sum} atom with weight $1$
and for the \lstinline{+} operator, we explicitly specify the two variables we want to add.
For example, the atom \lstinline|&sum{ 1,X : set(P,X), include(X) } = F|
assigns to variable \lstinline{F} the number of variables \lstinline{X}
that are contained in set \lstinline|P| and included in the solution.

As all these \lstinline{&sum} atoms appear in the heads of rules, they assign values to variables.
Note, however, that here we are not using the directional assignment operator \lstinline{=:}
but just normal equality.
In general, \flingo\ offers two distinct semantics:
One, which is evoked by using the operator \lstinline{=:}, only assigns a value to the variable on the right side,
if all variables on the left side are defined.
The second one, evoked by using normal equality, has no directionality and forces all variables occurring in the constraint to take on a value.
In that case, if both variables in the constraint \lstinline|&sum{X} = Y| were undefined,
the constraint would force them to take any value as long as they are equal (which could possibly result in an infinite amount of solutions).

In our case, the rules would behave the same under both semantics
because we are already checking for definedness of the variables on the left side
by means of the \lstinline{include(X)} atom.
This is possible due to the conditionality of the aggregate functions mentioned earlier.

We can solve a \coom\ instance using \flingo\ by adding the \lstinline{--solver flingo} (or \lstinline{-s flingo} for short) option
on the command-line.
By default, the \coomsuite\ uses \clingo\ as solver.
We do not show the command-line output here as it coincides with the one from \clingo.

\subsection{Adding user input}\label{sec:userinput}
\begin{lstlisting}[language=coom,label=lst:userinput:coom,caption={A \coom\ user input file for the \travelbike}]
add frame[0].bag[0]             |\label{lst:userinput:coom:bag}|
set color[0] = Yellow           |\label{lst:userinput:coom:color}|
set requestedVolume[0] = 200    |\label{lst:userinput:coom:volume}|
\end{lstlisting}
%
All the \coom\ language fragments described so far equip the user
with the ability to build up a \emph{configuration model} which can be seen as a blueprint for all
possible configurations.
Another important concept in configuration is that of \emph{user requirements} which allows the user to
specify knowledge that needs to be included in the current configuration~\citep{sotimasu98a}.
For example, the user could require the solution to include an optional component (e.g., the basket of a bike) or
to set the value for an attribute (e.g., defining a minimum number of storage space in the \travelbike).

The \coomsuite\ allows to perform this via a separate, so-called \emph{user input} file that can be passed along on the command-line.
For these means, the \coom\ language provides two keywords:
\lstinline{add} (Line~\ref{lst:userinput:coom:bag}) and
\lstinline{set} (Line~\ref{lst:userinput:coom:color}-\ref{lst:userinput:coom:volume})
which respectively, add an object to the configuration and set the value of an attribute.
For both, the variable identifiers from the refinement phase introduced in the previous section (cf.~Section~\ref{sec:extensions:asp:facts})
have to be used, however, without the quotation marks.
An example of such a user input for the \travelbike\ can be seen in Listing~\ref{lst:userinput:coom}.

During parsing and the refinement phase, the \coom\ user input
is converted into \lstinline{user_include/1} and \lstinline{user_value/2} predicates, respectively.
Then, before solving, the \coomsuite\ checks if the user input is valid with respect to the configuration model
and gives out warnings in case of inconsistencies.
Warnings are given out when:
\begin{enumerate}
  \item The referenced variable does not exist.
  \item A value is being set for a variable not corresponding to an attribute.
  \item The value is outside of the attribute domain.
\end{enumerate}
Note that as upper bounds of cardinalities are compiled away during the refinement process,
it is not necessary to check for exceeding maximal cardinalities.
Instead, this is implicitly handled by the first kind of warning.

Going back to our simplified \travelbike\ from Listing~\ref{lst:travelbike:coom},
note that the product model does not have a \lstinline{color} feature,
and thus the \coomsuite\ would give out a warning of the first case about Line~\ref{lst:userinput:coom:color} in Listing~\ref{lst:userinput:coom}.

Independently of any warnings, the user input is processed by the encoding in Listing~\ref{lst:encoding:userinput}.
However, before adding any user input to the final configuration, the encoding first verifies
for each \lstinline{user_include/1} and \lstinline{user_value/2} that the user input is consistent with the configuration model
(Lines~\ref{lst:encoding:user:consistent:add}-\ref{lst:encoding:user:consistent:discrete:2}).
We omit here the rules for integer domains as the check is similar as for discrete domains.
\begin{lstlisting}[language=clingos,label=lst:encoding:userinput,linerange={1-3,8-11},caption={Solving user input in \clingo}]
consistent(X)   :- user_include(X), type(X,_).  %%#(\label{lst:encoding:user:consistent:add}#)
consistent(X,V) :- user_value(X,V), type(X,T),  %%#(\label{lst:encoding:user:consistent:discrete:1}#)
                   discrete(T), domain(T,V).    %%#(\label{lst:encoding:user:consistent:discrete:2}#)

include(X) :- user_include(X), consistent(X).   %%#(\label{lst:encoding:user:add}#)
include(X) :- user_value(X,V), consistent(X,V). %%#(\label{lst:encoding:user:add:value}#)
value(X,V) :- user_value(X,V), consistent(X,V). %%#(\label{lst:encoding:user:set}#)
\end{lstlisting}

For each \lstinline{user_include/1} predicate we check that the object to be added exists.
For a \lstinline{user_value/2} predicate we check that the attribute exists and that the value is in the domain of the attribute.
Note that these checks only cover simple inconsistency cases and
it is still possible for the user to provide inconsistent input such as an invalid combination of attribute values (which might be harder to detect).

Subsequently, in Lines~\ref{lst:encoding:user:add}-\ref{lst:encoding:user:set}
the consistent user input is added to the configuration.
If the object in an \lstinline{user_include/1} predicate refers to an attribute this assures that a value will be set
for this attribute (but the user does not need to specify this value).
Any user input not consistent with the configuration model is ignored.

We can solve a \coom\ instance with user input by using the \lstinline{--user-input} option (or \lstinline{-u} for short)
and passing the path to the user input file as an argument.
In~Listing~\ref{lst:user:cli} we are running the simplified \travelbike\ example from before with the user input from Listing~\ref{lst:userinput:coom}.
The \coomsuite\ prints out a warning for the \lstinline[language=coom]{set color[0] = Yellow} directive, as the variable is not part of the configuration model
but proceeds to solve the configuration problem, ignoring the incosistent user input.
%
\begin{lstlisting}[language=shell,label=lst:user:cli,caption={Solving the simplified \travelbike\ example together with user input}]
$ coomsuite solve travel-bike-simplified.coom -u user-input.coom -o coom
WARNING:  - Invalid user input.
Variable root.color[0] does not exist.
COOM Suite version 0.1
Reading from refined-input.lp
Solving...
Answer: 1
carrier[0]  frame[0]
requestedVolume[0] = 200    totalVolume[0]     = 200
carrier[0].bag[0] = "B50"   carrier[0].bag[0].volume[0] = 50
carrier[0].bag[1] = "B50"   carrier[0].bag[1].volume[0] = 50
carrier[0].bag[2] = "B50"   carrier[0].bag[2].volume[0] = 50
frame[0].bag[0]   = "B50"   frame[0].bag[0].volume[0]   = 50
SATISFIABLE
\end{lstlisting}

\subsection{Reasoning with unbounded cardinalities}\label{sec:open}

In all earlier \coom\ language fragments a lower and upper bound for feature cardinalities had to be explicitly specified
(or the cardinality defaulted to $1..1$ if no bounds were given).
In practice, however, when modeling a configuration problem, it is often not possible to determine the exact (or even approximate) number of objects needed in advance~\citep{tafasc16a}.
In this section we extend \poom\ and \xoom\ to support unbounded cardinalities
and call these \openpoom\ and \openxoom, respectively.
Then, we present a simple modification of our existing workflow to solve problems of such kind.

Listing~\ref{lst:cargobike:coom} shows the \cargobike,
a slightly modified version of the \travelbike\ example from Listing~\ref{lst:travelbike:coom} 
such that the number of possible bags is now unknown.
%
\begin{lstlisting}[language=coom,label=lst:cargobike:coom,caption={Representation of the \cargobike\ example in \coom}]
product {
            num  0-200   totalVolume
            num  0-200   requestedVolume
    0..*    Bag          bags
}
enumeration Bag {
    attribute num volume

    small   = ( 12 )
    large   = ( 20 )
}
behavior {
    require sum(bags.volume) = totalVolume

    require totalVolume >= requestedVolume
}
\end{lstlisting}
%
As in the \travelbike, the \cargobike\ has two numeric features \lstinline{totalVolume} and \lstinline{requestedVolume}
where the former is computed by summing up the volume of all bags and the latter specifies a minimum for the former (computed) value.
However, different from before, the feature \lstinline{bags} has cardinality \lstinline{0..*},
which stands for zero or more bags and is unbounded above.
This can not be solved natively with the encoding presented in Section~\ref{sec:extensions:asp}.

\begin{figure}
  \centering
  \includegraphics[width=\textwidth]{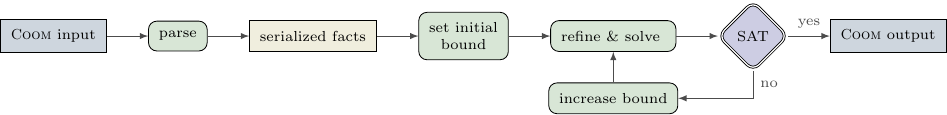}
  \caption{Workflow for Incremental Bounds algorithm}
  \label{fig:incremental-bounds}
\end{figure}
%
As a simple workaround, we set a maximum bound during the refinement phase and check whether this leads to a solution.
The \coomsuite\ currently offers two search strategies to determine this maximum bound:
A linear and an exponential one~\citep{frscel21a}.
We call this approach \emph{Incremental Bounds} and proceed in the following manner (cf.~Figure~\ref{fig:incremental-bounds}):
\begin{enumerate}
  \item Parse the \coom\ model into the serialized ASP fact format.
  \item Set the initial maximum upper bound to $n$ (typically $n=1$).
  \item Refine and solve the configuration problem.
  \item If no solution is found, determine a new maximum bound according to the chosen strategy and repeat from step 3.
\end{enumerate}
While the linear search strategy simply increases the maximum bound by a chosen value $k$ (typically $k=1$),
the exponential one is inspired by~\cite{frscel21a}
and first doubles the current bound until finding a solution and then converges to the optimal bound.

While this approach is very simple, its advantages are that it is general (works with any combination of open cardinalities) and
does not require any changes to the ASP encoding.
Further, at least in an approximate way,
it provides a built-in minimization on the number of parts.
However, it does not (yet) make use of \clingo's multi-shot solving capabilities,
and therefore grounding has to performed again on every iteration.

To solve a \coom\ instance with incremental bounds, we can use the \lstinline{--bounds} option
to specify the desired search strategy
as shown in Listing~\ref{lst:cargobike:cli}.
Here, we have chosen linear search and set the value of \lstinline{requestedVolume} to $60$ via a user input file.
As the maximum volume of a single bag is $20$, the iteration is not able to find a solution until the maximum bound is set to $3$.
%
\begin{lstlisting}[language=shell,label=lst:cargobike:cli,caption={Solving the \cargobike\ example with the incremental bounds option}]
$ coomsuite cargo-bike.coom -u input.coom --bounds=linear -o coom

Solving with max_bound = 1
UNSATISFIABLE

Solving with max_bound = 2
UNSATISFIABLE

Solving with max_bound = 3
COOM Suite version 0.1
Reading from refined-input.lp
Solving...
Answer: 1
bags[0]
bags[1]
bags[2]
requestedVolume[0] = 60
totalVolume[0]     = 60

bags[0].size[0] = "large"
bags[0].size[0].volume[0] = 20  bags[0].size[0].weight[0] = 25

bags[1].size[0] = "large"
bags[1].size[0].volume[0] = 20  bags[1].size[0].weight[0] = 25

bags[2].size[0] = "large"
bags[2].size[0].volume[0] = 20  bags[2].size[0].weight[0] = 25

SATISFIABLE
\end{lstlisting}

\section{Solving \coom\ interactively}\label{sec:ui}

While in the previous sections the workflow assumed an input from the user only at the beginning
of the process, in practice, users typically want to interact with the system continuously,
e.g., to experiment with different settings, explore the solution space or just to debug the configuration model.
This is especially true for complex configuration problems where the number of solutions is large
and difficult to envision by just looking at an input model (whether in textual or graphical form).
In that sense, a user interface (UI) for solving \coom\ interactively is a valuable
addition to the \coomsuite\ workbench.

For that purpose, we next present a prototypical UI
which is generated and driven by ASP, more precisely by the \clinguin\ system~\citep{behasc24a}.
This system uses a simple design with dedicated predicates to define a UI and the behavior of user-triggered events,
thereby greatly facilitating the specification of continuous user interactions with an ASP system.
For details about the syntax and functionality of \clinguin\ we refer the reader to \citep{behasc24a}.

The UI presented in this section uses the \xoom\ encoding presented in Section~\ref{sec:extensions:asp}
plus an additional \clinguin\ encoding defining its layout, style and functionality.
Since integration with constraint systems such as \flingo\ is currently limited in \clinguin,
the UI only works with \clingo.

User input as introduced in Section~\ref{sec:userinput} is no longer needed.
Instead, the user can interactively make choices to create a configuration solution.
For this, we use assumptions, which can be interpreted as integrity constraints that force
the encoding to entail the provided atom.
As before with the user input,
these assumptions can be used to set values to attributes or to force the inclusion of a part in the final configuration.
Moreover, the UI provides additional functionality for the user
to browse solutions, download the current \coom\ solution and obtain
basic explanations of why a selection is not valid.

We start by showcasing a simple interaction of the UI using the \travelbike\ example
by presenting snippets of the UI encoding together with screenshots of the possible user interactions.
The full UI encoding can be found in \citep{coomsuite}.
Then, we proceed to show the explanation features of the UI and how they are implemented in \clinguin.

Furthermore, to better demonstrate the UI capabilities, we do not use the simplified version of the \travelbike\ from above but
instead use the complete example.
The main differences are that the \lstinline{Wheel} and \lstinline{Bag} enumerations
additionally have a \lstinline{weight} attribute contributing to the \lstinline{totalWeight} of the bike
and which can be constrained by a user-set value \lstinline{maxWeight}.
There are also additional \lstinline{feature}[s] such as the \lstinline{color} of the bike
and the \lstinline{material} of the \lstinline{Bag}.
Lastly, the \travelbike\ contains some additional constraints among which we only highlight the
conditional requirement in Listing~\ref{lst:travelbike:condreq}
stating that the \lstinline{color} \lstinline{Red} implies a \lstinline{frontWheel} of size \lstinline{20}.
%
\begin{lstlisting}[language=coom,label={lst:travelbike:condreq},linerange=73-74,caption={Conditional requirement of the complete \travelbike\ example}]
condition color = Red
require frontWheel.size = 20
\end{lstlisting}
%

\paragraph{Attributes and values}

\begin{figure}
  \centering
  \includegraphics[width=.9\textwidth,trim={0 13.5cm 0 0}]{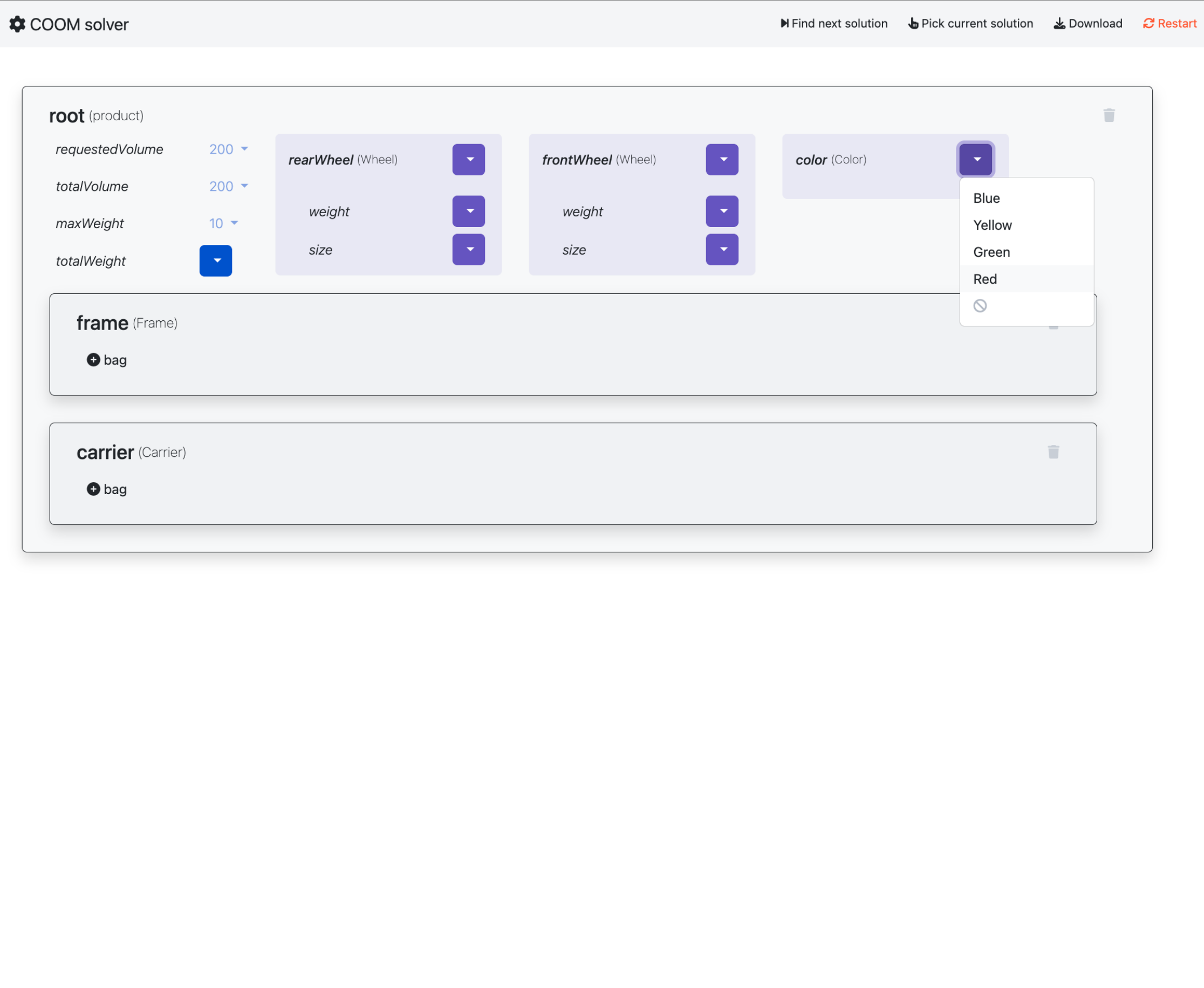}
  \caption{Initial state of UI}
  \label{fig:ui:initial}
\end{figure}
\begin{figure}
  \centering
  \includegraphics[width=.9\textwidth,trim={0 13.5cm 0 0}]{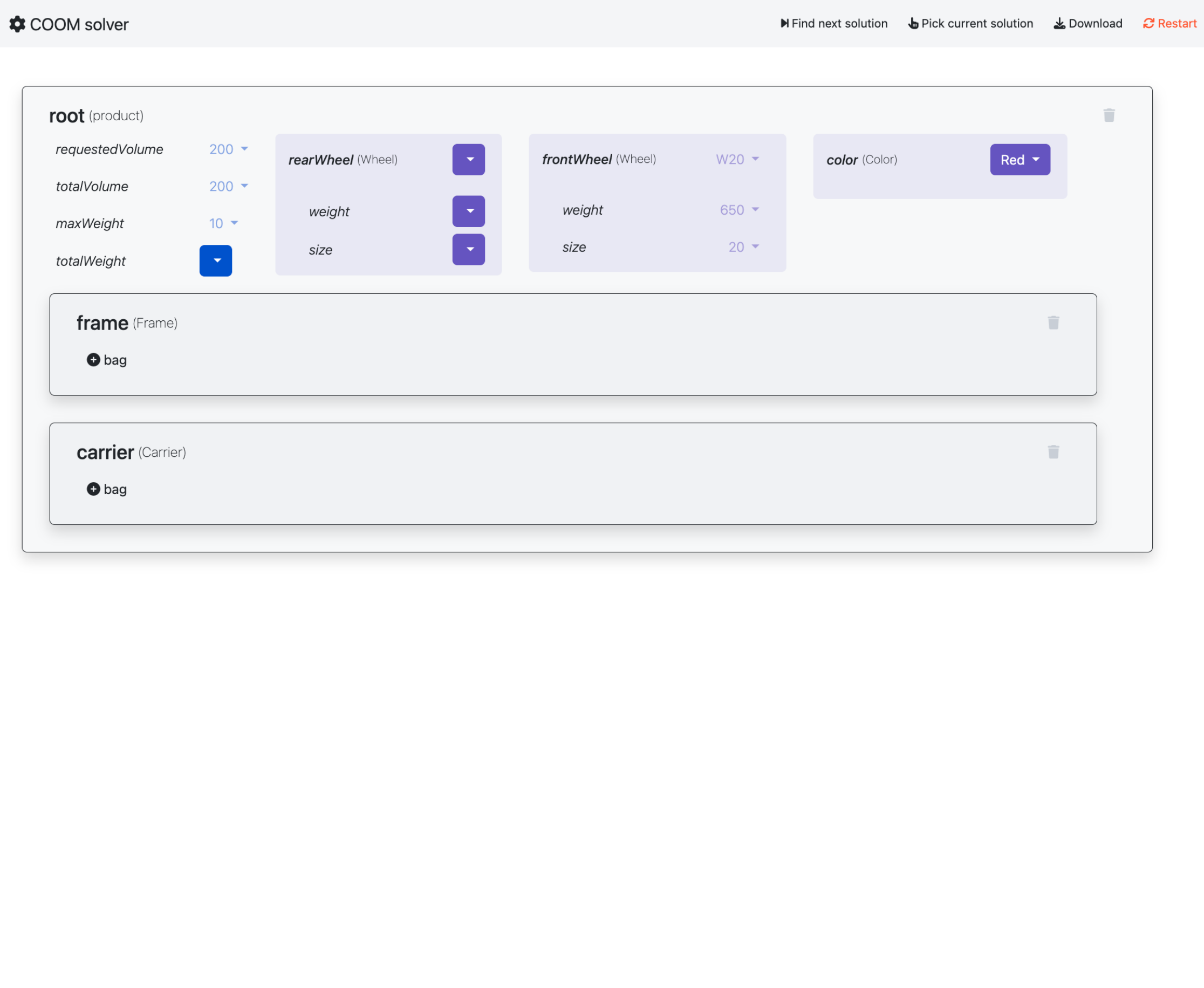}
  \caption{UI after selecting the \lstinline{color} \lstinline{Red}}
  \label{fig:ui:infer}
\end{figure}

%
Figure~\ref{fig:ui:initial} displays the initial state of the UI upon loading
with an opened dropdown menu showing the available options for the \lstinline{color} of the bike.
Upon selecting \lstinline{Red} as the \lstinline{color}, the UI updates to show any inferred values
resulting from this choice.
In this case, the constraint from Listing~\ref{lst:travelbike:condreq} causes a \lstinline{frontWheel} of size \lstinline{20}
to be inferred and this is updated accordingly in the UI (cf.~Figure~\ref{fig:ui:infer}).
%
%
We now explain how this is represented in the \clinguin\ encoding.
For the sake of brevity, we only show exemplary snippets of the encoding relevant to attributes and values.
Note that while the UI is built to represent the \coom\ input language,
the encoding is using the predicates (and naming conventions) of the ASP fact format (cf.~Section~\ref{sec:extensions:asp}).

The first two Lines of Listing~\ref{lst:ui:attr:initial} create auxiliary predicates for the two types of attributes in our fact format.
Subsequently, these are used in Lines~\ref{lst:encoding:ui:iattr} and \ref{lst:encoding:ui:iattr-browse}
to create the predicate \lstinline|i_attr(X,AT)| stating that attribute variable $X$ of type $AT$ is included (in the solution).
%
\begin{lstlisting}[language=clingos,label=lst:ui:attr:initial,linerange=1-6,caption={Definition of included attributes and their values in the UI}]
attr_type(X,discrete) :- type(X,T), discrete(T).                                %%#(\label{lst:encoding:ui:attrd}#)
attr_type(X,integer)  :- type(X,T), integer(T).                                 %%#(\label{lst:encoding:ui:attri}#)
i_attr(X,AT) :- attr_type(X,AT), _all(include(X)).                              %%#(\label{lst:encoding:ui:iattr}#)
i_attr(X,AT) :- attr_type(X,AT),      include(X), _clinguin_browsing.           %%#(\label{lst:encoding:ui:iattr-browse}#)
i_value(X,V) :- i_attr(X,_), _all(value(X,V)).                                  %%#(\label{lst:encoding:ui:ival}#)
i_value(X,V) :- i_attr(X,_),      value(X,V), _clinguin_browsing.               %%#(\label{lst:encoding:ui:ival-browse}#)
\end{lstlisting}
%
Here, Line~\ref{lst:encoding:ui:iattr} checks whether an attribute variable is included in all solutions
using the dedicated \clinguin\ predicate \lstinline|_all/1|.
This allows us to show only relevant attributes in the UI.
In the two screenshots, for instance, we do not see any attributes for the possible \lstinline{bag}[s]
as they are not included in all solutions (thus not mandatory).
Line~\ref{lst:encoding:ui:iattr-browse}, on the other hand, checks what attribute variables are included
while browsing through solutions.
Similarly, Lines~\ref{lst:encoding:ui:ival} and \ref{lst:encoding:ui:ival-browse}
define the (attribute) values to be shown via predicate \lstinline|i_value(X,V)|.

With these auxiliary predicates, the first rule in Listing~\ref{lst:ui:attr:dropdown}
creates a dropdown menu on the UI for each attribute and
the second rule adds the text for this dropdown menu
when a value exists.
We can see this when comparing the first and second screenshot in Figures~\ref{fig:ui:initial} and \ref{fig:ui:infer}
where the text in the dropdown menus of the \lstinline{color} and \lstinline{frontWheel} enumerations changes
accordingly.
\begin{lstlisting}[language=clingos,label=lst:ui:attr:dropdown,linerange=9-11,caption={Definition of UI dropdown menus}]
elem(dd(X), dropdown_menu,
            attr_container(X)) :- i_attr(X,_).                                   %%#(\label{lst:encoding:ui:elemdd}#)
attr(dd(X),selected,V)         :- i_value(X,V).                                  %%#(\label{lst:encoding:ui:selected}#)
\end{lstlisting}
%

The style of the dropdown menus is defined in Listing~\ref{lst:ui:attr:style}
where the first two rules set the style for inferred values.
The first rules add the light text with low opacity for values that are inferred but not (yet) selected,
e.g., as for the \lstinline{frontWheel}.
Furthermore, the second rule removes the border of the dropdown menus of such attributes to better distinguish them.

Next, the third and fourth rule add the class \lstinline|"btn-secondary"|, which is the purple button style,
for mandatory discrete attributes (stemming from enumerations in \coom) for which either a value has been selected or there are multiple options left.
Note that for numeric values (not shown here) this works similarly but with the class \lstinline|"btn-primary"|, which is the blue button style.
Selecting values in \clinguin\ is done via adding assumptions to \clingo\ for the corresponding atom
and the predicate \lstinline{_clinguin_assume/2} is added by \clinguin\ for every atom that has been assumed to be true or false.
%
\begin{lstlisting}[language=clingos,label=lst:ui:attr:style,linerange=13-21,caption={Definition of UI style for attributes}]
attr(dd(X),class,("fw-light";
                  "opacity-50"))  :- i_value(X,V), _all(value(X,V)),
                                     not _clinguin_assume(value(X,V),true).
attr(dd(X),class,"border-0")      :- i_attr(X,discrete), i_value(X,_),
                                     not _clinguin_assume(value(X,_),true).
attr(dd(X),class,"btn-secondary") :- i_attr(X,discrete),
                                     not i_value(X,_).
attr(dd(X),class,"btn-secondary") :- i_attr(X,discrete),
                                     clinguin_assume(value(X,_),true).
\end{lstlisting}
%

The rules in Listing~\ref{lst:ui:attr:valid} are responsible for providing
the options of the dropdown menus that are still valid.
The first rule creates an auxiliary predicate with all possible values which
is subsequently used in the last three rules to create one dropdown menu item for each such value.
\begin{lstlisting}[language=clingos,label=lst:ui:attr:valid,linerange=27-32,caption={Definition of possible values for UI dropdown menus}]
v_option((X,V),valid) :- i_attr(X,_), _any(value(X,V)),
                         not _clinguin_assume(value(X,_),true).
elem(ddi(X,V), dropdown_menu_item, dd(X))  :- v_option((X,V),_).
attr(ddi(X,V), label, V)                   :- v_option((X,V),_).
when(ddi(X,V), click, call,
               add_assumption(value(X,V))) :- v_option((X,V),_).   %%#(\label{lst:encoding:ui:click}#)
\end{lstlisting}
%
Upon clicking on one of these values, a call is made to the solver to add the assumption of the corresponding value assignment,
thus forcing the entailment of the given value and restricting the possible solutions (cf.~Line~\ref{lst:encoding:ui:click}).
The dedicated \clinguin\ predicate \lstinline{_any/1} is the counterpart for \lstinline{_all/1}
and specifies that an atom is contained in at least one stable model.

\begin{figure}
  \centering
  \includegraphics[scale=0.26]{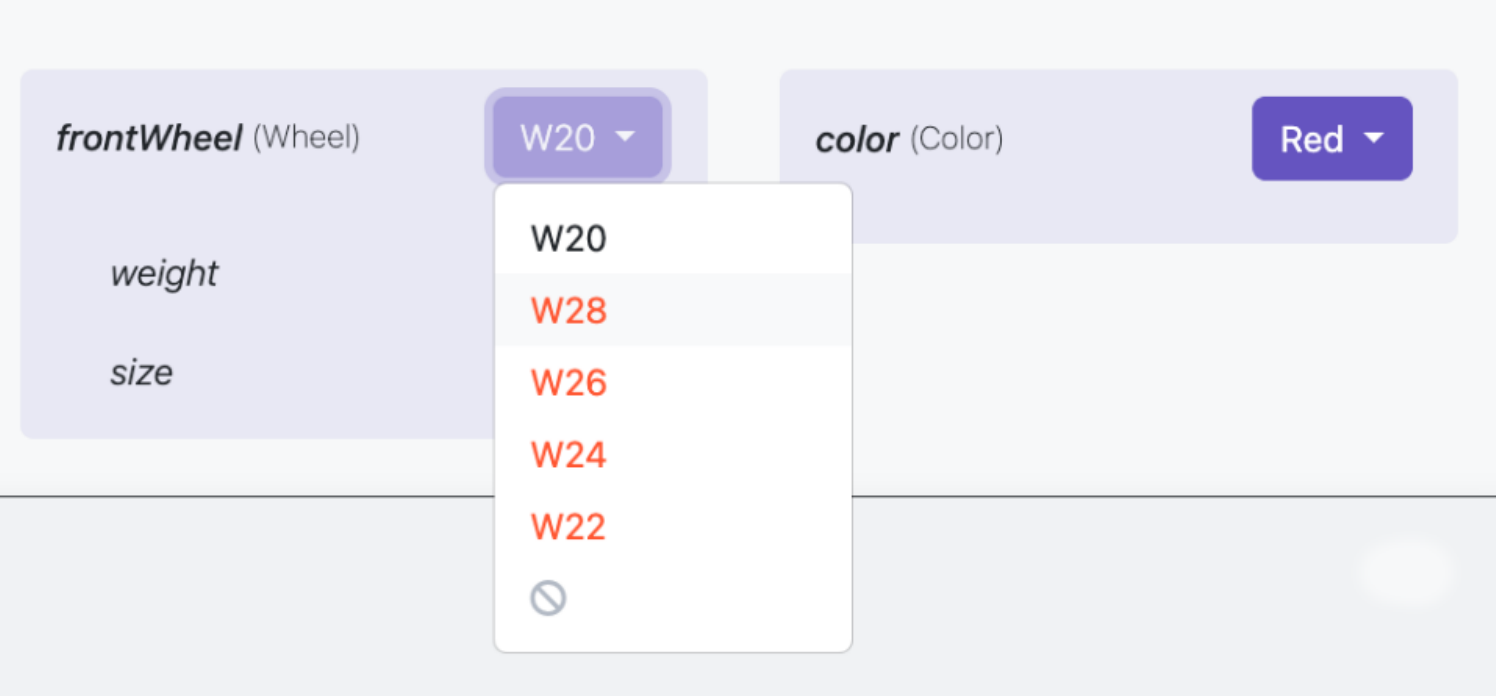}
  \includegraphics[scale=0.26]{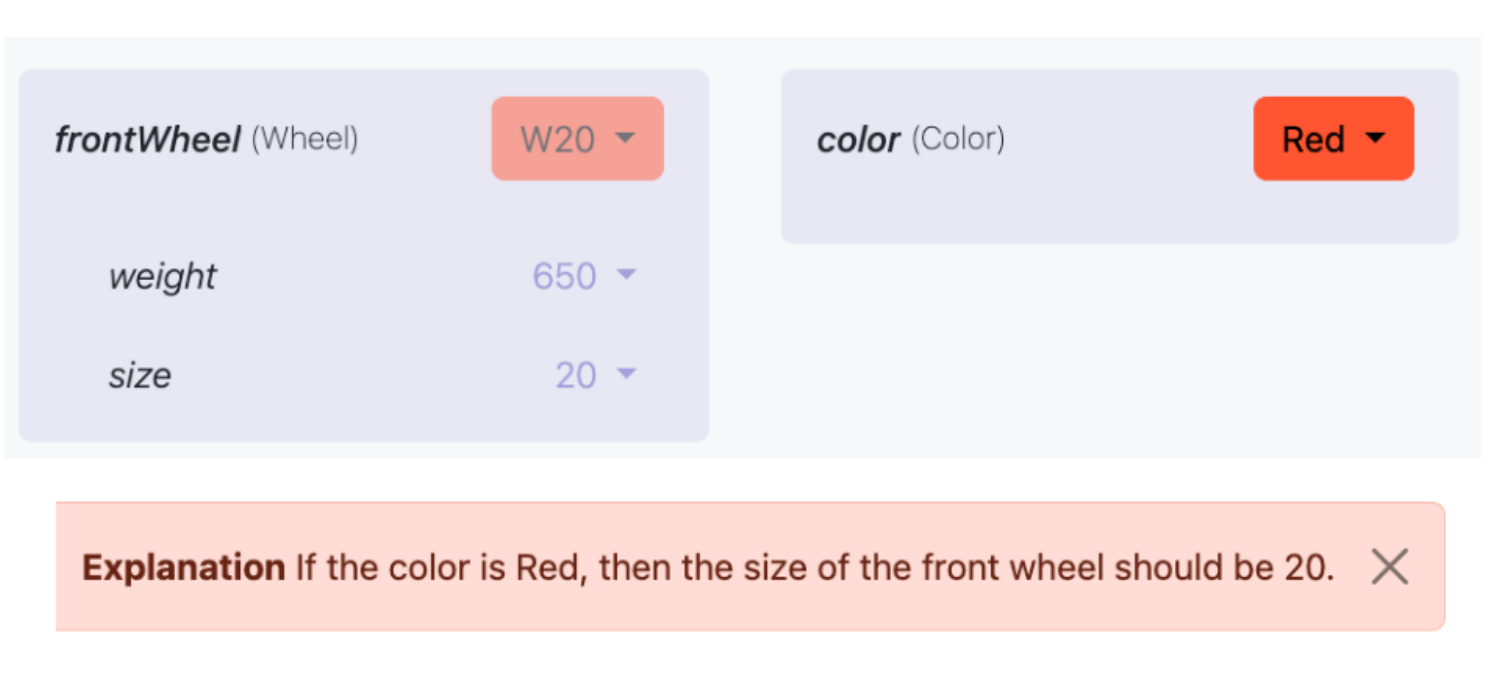}
  \caption{Explanations in the UI after selecting an invalid value}
  \label{fig:ui:explanation}
\end{figure}

\paragraph{Explanations}
The two cropped screenshots in Figure~\ref{fig:ui:explanation} show the explanation features of the UI.
Recall the constraint in Listing~\ref{lst:travelbike:condreq} stating that the \lstinline{color} \lstinline{Red}
implies a \lstinline{frontWheel} of size \lstinline{20} (which belongs to option \lstinline{W20})
and that the UI inferred the latter value upon selecting \lstinline{Red} as the \lstinline{color}.
Accordingly, when we open the dropdown menu of the \lstinline{frontWheel}, we see the other (invalid) options in red
but it is still possible to click on them.
When selecting an invalid value, however, the affected values of the \lstinline{frontWheel} and the \lstinline{color}
turn red and the UI shows an explanation of why the newly selected value is not valid.
We now explain how this works inside \clinguin.
%
%

Our approach to explanations for invalid values follows \citep{behasc24a}.
In a nutshell, this means that we use a specialized \clinguin\ backend which provides
a minimal unsatisfiable set (MUS) of assumptions whenever the encoding is unsatisfiable.
In this way, we allow the user to select invalid values, and thus to let the problem become unsatisfiable,
while subsequently using the information about faulty assumptions
to highlight previously selected values in red.
For our approach, this has the limitation that it only works for invalid selections of attribute values
but not for the inclusion of an invalid part.

Additionally, we leverage the definition of constraints in the refined ASP instance
to provide the user with a natural language explanation of why a certain value is invalid.
To do this, we indicate to the explanation backend
that the atoms of predicate \lstinline|constraint/2| should be treated as assumptions, and thus
be included as part of the reasons for unsatifiablity.
Moreover, we extend the ASP fact format for the configuration
to include the predicate \lstinline|explanation/2|
to store natural language explanations for Boolean constraints.
For instance, the constraint
\lstinline{!root.color[0]=Red||root.frontWheel[0].size[0]=20}
has a natural language explanation:
\textit{``If the color is red, then the size of the front wheel should be 20."}

Thanks to the generality of Boolean operators,
we were able to automatically generate these explanations for all Boolean constraints
using a Large Language Model (LLM) by prompting it with two simple examples.
However, they can also be provided by the user.
In fact, \coom\ offers a dedicated keyword for this which we plan
to integrate in the workflow of the \coomsuite\ in the future.
While it is desirable to have these kinds of explanations for all constraints,
we limit ourselves to Boolean constraints for now
and leave explanations for table constraints as future work as their treatment is more complex.

A snippet of the corresponding UI encoding is shown in Listing~\ref{lst:ui:explain}.
%
\begin{lstlisting}[language=clingos,label=lst:ui:explain,linerange={1-8},caption={UI explanation encoding}]
attr(ddi(X,V),class,
              ("text-danger")) :- i_attr(X,_), type(X,T),
                                  domain(T,V), not _any(value(X,V)).

attr(dd(X), class, ("btn-danger")) :- _clinguin_mus(value(X,V)).
elem(m(C), message, window) :- _clinguin_mus(constraint(C,"boolean")).
attr(m(C), message, M)      :- _clinguin_mus(constraint(C,"boolean")),
                               explanation(C,M).
\end{lstlisting}
%

The first rule sets the red text style for the invalid dropdown menu items
which belong to domain values not contained in any solution.
The other three rules are utilizing the dedicated \clinguin\ predicate \lstinline|_clinguin_mus/1|
which provides information about the MUS.
While the first rule sets the red button style for all values contained in the MUS,
the second and third rule provide the (natural language) explanation inside a separate window (cf.~Figure~\ref{fig:ui:explanation}).

\begin{figure}
  \centering
  \includegraphics[width=.9\textwidth,trim={0 6.5cm 0 0}]{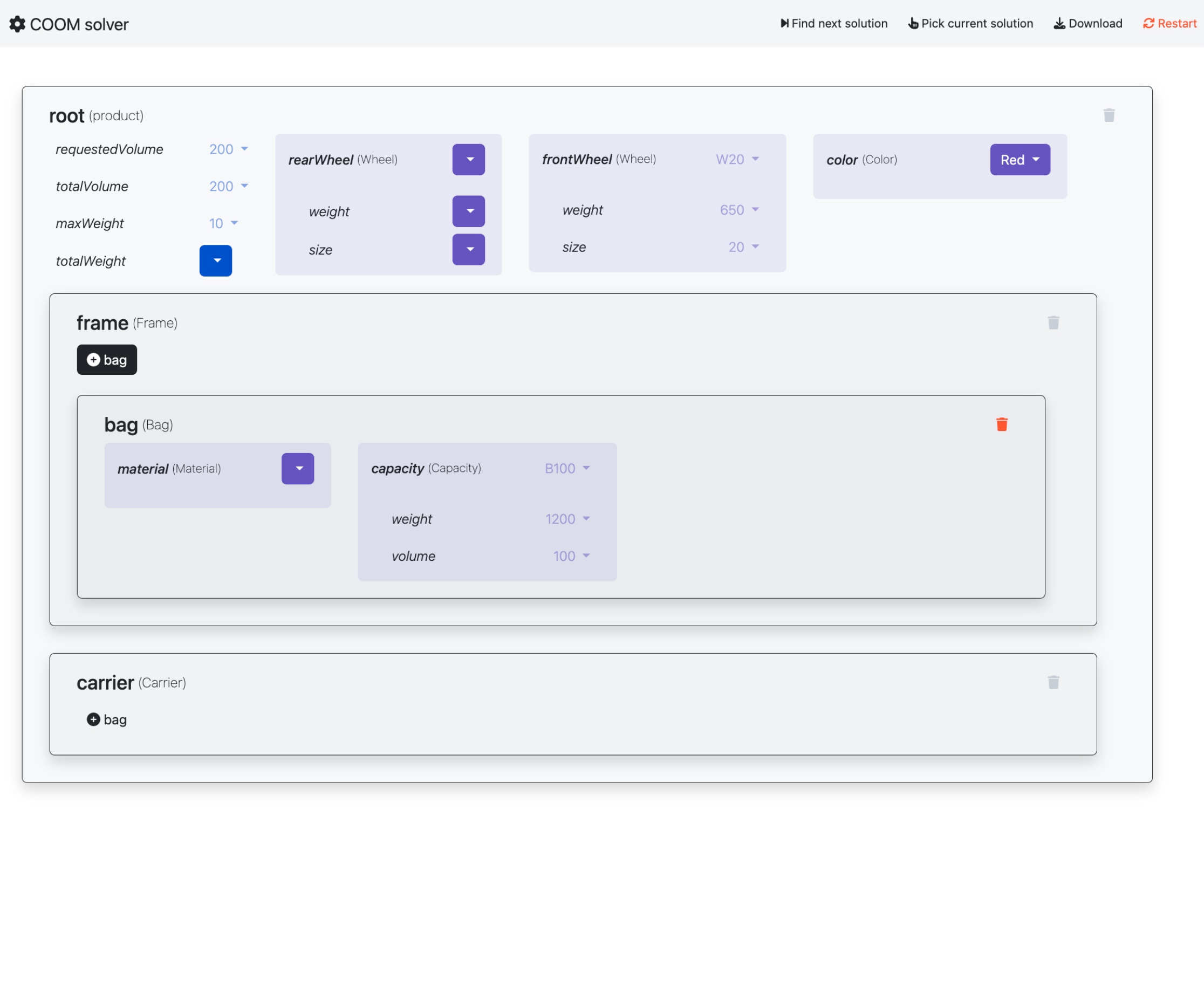}
  \caption{Adding a \lstinline{bag} to the solution}
  \label{fig:ui:adding}
\end{figure}

Finally, leaving behind the topic of explanations,
the last screenshot shows the addition of a \lstinline{bag} to the configuration (cf.~Figure~\ref{fig:ui:adding})
by means of clicking on the small icon containing a $+$.
Since it was optional to add this part, it includes a red button to remove it.
For the sake of brevity, we do not go into further detail here.
%
%
As next steps, the user might decide to browse the different solutions, pick one and continue modifying it,
or download it in the format of a \coom\ solution.
The interested reader can try this out by following the instructions in~\citep{coomsuite}.
Currently, this requires running \clinguin\ separately but for future versions of the \coomsuite\
a direct integration is planned.

\section{The \coomsuite\ Workbench}\label{sec:benchmarks}

The \coomsuite~\citep{coomsuite} is intended to serve as a workbench for experimentation with
industrial-scale product configuration problems.
While the included benchmark collection can be utilized with other paradigms,
its current infrastructure is primarily geared towards ASP.
Specifically, the \coomsuite\ is available as a \python\ package, installable via \pip.
It includes a (customizable) ANTLR~v4 parser to convert \coom\ specifications into facts,
along with an ASP encoding to harmonize the fact format with the chosen configuration encoding.
The current distribution includes
four scalable benchmark series,
a single ASP encoding covering all essential configuration concepts needed for the three \coom\ language fragments,
as well as an additional one for the hybrid solver \flingo~\citep{flingo}.

\begin{table}[ht]
  \centering
  \begin{tabular}{ |c|c|c|c|c| }
    \cline{1-5}
    Benchmark set    & Language & Description              & Scalable factor        & \# Instances \\
    \cline{1-5}
    \textit{Core}    & \core    & Random table constraints & \#Attributes, \#Values & 45           \\
    \citybikefleet   & \poom    & Fleet of \citybikes      & \#Bikes                & 15           \\
    \travelbikefleet & \xoom    & Fleet of \travelbikes    & \#Bikes                & 15           \\
    \cline{1-5}
  \end{tabular}
  \caption{Benchmark sets of the \coomsuite}
  \label{tab:benchmarks}
\end{table}

For each of the three \coom\ language fragments, we include a benchmark set in the \coomsuite\ detailed in Table~\ref{tab:benchmarks}.
The two language extensions \openpoom\ and \openxoom\ currently have no dedicated benchmark sets.
Additionally, the \coomsuite\ includes the benchmark set of a \restaurant\ corresponding to the \xoom\ language,
however, with much simpler numerical calculations.
Here, the aim is to configure the assignment of a given number of chairs to tables of different sizes.
It makes use of partonomy as well as simple numeric constraints and the scalable factor is the total number of chairs needed.
We decided to not show the results of the \restaurant\ domain here,
as they were similar to the \textit{Core} and \citybikefleet\ domains.

\begin{figure}[ht]
  \includegraphics[width=.8\linewidth]{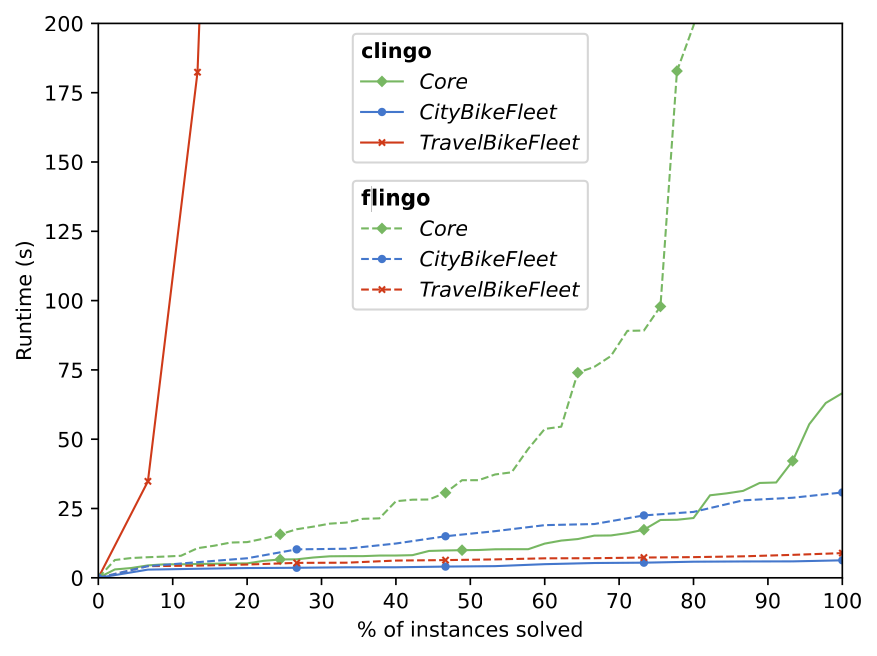}
  \caption{Runtimes for benchmarks of Table~\ref{tab:benchmarks}}
  \label{fig:results:all}
\end{figure}
To provide a glimpse into the usage of the \coomsuite,
we conducted sample benchmarks comparing the standard ASP encoding (Section~\ref{sec:extensions:asp:encoding})
with that designed for \flingo\ (Section~\ref{sec:extensions:asp:flingo}), which utilizes integer variables.
We ran all instances of the benchmarks in Table~\ref{tab:benchmarks} on a compute cluster with Intel Xeon E5-2650v4@2.9GHz~CPUs with 64GB of memory running
Debian Linux~10.\footnote{\url{https://www.cs.uni-potsdam.de/bs/research/labs.html#hardware}}
We used a timeout of 300 seconds and limited the memory to 16GB per instance.
Figure~\ref{fig:results:all} shows the runtimes of finding one stable model for both solvers \clingo\ and \flingo.
On the two non-numeric domains (\textit{Core} and \citybikefleet) \clingo\ performs better than \flingo.
While \clingo\ is able to solve most instances within a couple seconds (and takes up to 60 seconds for bigger instances),
\flingo\ times out on some of the instances of the \textit{Core} domain.
In general the runtimes of \flingo\ are about one order of magnitude higher than those of \clingo.
However, for the \travelbikefleet, \flingo\ clearly outperforms \clingo\
due to the problem of the latter with large numeric ranges.
In contrast, \flingo\ can handle these effortlessly due to its native handling of integer variables.
Note that when \clingo\ times out it is usually during grounding,
while for \flingo\ this usually happens during its internal preprocessing which uses \clingo's \python\ API.

\section{Discussion}\label{sec:discussion}

Researchers often lack access to industrial-scale examples to evaluate their work due to confidentiality or limited
public availability.
To address this challenge in product configuration,
we introduce the \coomsuite,
a workbench offering a curated collection of product model benchmarks and tools for converting them to ASP.
These benchmarks, derived from industrial contexts, reflect the key challenges of product configuration.
The workflow includes a refinement step that separates the ASP representation of the \coom\ input from the
representation tailored to specific ASP encodings or systems.
We highlighted the design of a series of such ASP encodings that handle increasingly complex \coom\ models.

Our work provides not only the first publicly available ASP implementation of \coom\
but also (indirectly) establishes first semantic underpinnings for \coom.
However, this is just the starting point.
Future challenges include addressing unimplemented \coom\ features
such as optimization and explanations,
as well applying and further developing alternative ASP configuration encodings~\citep{faryscsh15a,gescer19a,ruscst23a}
within this uniform industrial-scale setting.
Concrete improvements include a normalization of constraints during parsing which enables the reutilization of information during solving,
studying a more native representation of \coom\ for \flingo,
and the development of a truly incremental approach for unbounded cardinalities using \clingo's multi-shot solving capabilities.
While the current approach to the latter provides a simple and general baseline solution,
a multi-shot approach will most likely provide a much better performance.
However, this requires greater modifications to the ASP encoding
and further study on how to handle multiple, especially nested, open cardinalities effectively
as well as an extension of the benchmark sets for the sake of an exhaustive evaluation.

As what concerns transparency and user integration,
we can leverage ASP-driven visualization~\citep{hasascst22a} and
enhance the capabilities for interactive exploration of the configuration space
by means of further conceptual and technical development of the UI.

We also aim to expand the model benchmarks in the \coomsuite\ with more domains inspired by real-world applications,
creating a comprehensive and challenging workbench for advancing ASP development and beyond.

\end{document}

%% file: macro.tex
\providecommand{\Underscore}{\textunderscore}

\lstdefinelanguage{clingo}{%
basicstyle=\ttfamily,%
keywordstyle=[1]\bfseries,%
keywordstyle=[2]\bfseries,%
keywordstyle=[3]\bfseries,%
showstringspaces=false,%
literate={_}{\Underscore}1 {\%\%}{}0,%
escapeinside={\#(}{\#)},%
alsoletter={\#,\&},%
keywords=[1]{not,from,import,def,if,else,elif,return,while,break,and,or,for,in,del,and,class,with,as,is,yield,async},%
keywords=[2]{\#const,\#show,\#minimize,\#base,\#theory,\#count,\#external,\#program,\#script,\#end,\#heuristic,\#edge,\#project,\#show,\#sum},%
morecomment=[l]{\#\ },%
morecomment=[l]{\%\ },%
morestring=[b]",%
stringstyle={\itshape},%
commentstyle={\color{darkgray}}%
}

\lstdefinelanguage{clingcon}[]{clingo}{morekeywords={&dom,&sum,&nsum,&diff,&disjoint,&distinct,&minimize,&maximize,&show}}
\lstdefinelanguage{flingo}[]{clingo}{morekeywords={&sum,&sus,&in,&df,&min,&max,&show}}
\lstdefinelanguage{clingodl}[]{clingo}{morekeywords={&diff}}

\lstdefinelanguage{python}{%
    basicstyle=\ttfamily,%
    keywordstyle=[1]\bfseries,%
    showstringspaces=false,%
    literate={_}{\Underscore}{1},%
    escapeinside={\#(}{\#)},%
    alsoletter={\#,\&},%
    keywords=[1]{not,from,import,def,if,else,elif,return,while,break,and,or,for,in,del,and,class,with,as,is,yield,async},%
    morecomment=[l]{\#\ },%
    morestring=[b]",%
    stringstyle={\itshape},%
    commentstyle={\color{darkgray}}%
}

\lstdefinelanguage{clingos}{%
    language=clingo,%
    basicstyle=\small\ttfamily,%
}
\lstdefinelanguage{flingos}{%
    language=flingo,%
    basicstyle=\small\ttfamily,%
}
\lstdefinelanguage{coom}{
    basicstyle=\small\ttfamily,
    keywordstyle=[1]\bfseries,
    keepspaces=true,
    morecomment=[l]{\//\ },
    tabsize=2,
    keywords=[1]{product,structure,enumeration,attribute,num,behavior,require,condition,combinations,allow,imply,add,set},
    sensitive=false,
    escapechar=|,
}
\lstdefinelanguage{shell}{frame=single,basicstyle=\footnotesize\ttfamily,numbers=none,escapeinside={|}{|},xrightmargin=2\parindent }

\providecommand{\sysfont}{\textit}
\newcommand{\clingo}{\sysfont{clingo}}
\newcommand{\flingo}{\sysfont{flingo}}
\newcommand{\lctocasp}{\sysfont{lc2casp}}
\newcommand{\clingcon}{\sysfont{clingcon}}
\newcommand{\clinguin}{\sysfont{clinguin}}

\newcommand{\python}{Python}
\newcommand{\pip}{pip}

\newcommand{\CoomFont}[1]{\textsc{#1}}
\newcommand{\coom}{\CoomFont{Coom}}
\newcommand{\Coom}{\coom}
\newcommand{\core}{\CoomFont{\coom{Core}}}

\newcommand{\poom}{\CoomFont{\coom[p]}}

\newcommand{\xoom}{\CoomFont{\Coom[x]}}

\newcommand{\openpoom}{\CoomFont{\coom[p*]}}
\newcommand{\openxoom}{\CoomFont{\coom[x*]}}

\newcommand{\coomsuite}{\CoomFont{CoomSuite}}

\newcommand{\ExFont}[1]{\textit{#1}}
\newcommand{\kidsbike}{\ExFont{KidsBike}}
\newcommand{\citybike}{\ExFont{CityBike}}
\newcommand{\travelbike}{\ExFont{TravelBike}}
\newcommand{\cargobike}{\ExFont{CargoBike}}

\newcommand{\citybikes}{\ExFont{CityBikes}}
\newcommand{\travelbikes}{\ExFont{TravelBikes}}

\newcommand{\citybikefleet}{\ExFont{CityBikeFleet}}
\newcommand{\travelbikefleet}{\ExFont{TravelBikeFleet}}
\newcommand{\restaurant}{\ExFont{Restaurant}}

\newcommand{\eqdef}{%
    \mathrel{\vbox{\offinterlineskip\ialign{%
                \hfil##\hfil\cr%
                $\scriptscriptstyle\mathrm{def}$\cr%
                \noalign{\kern1pt}%
                $=$\cr%
                \noalign{\kern-0.1pt}%
            }}}}